\documentclass[a4paper,USenglish,cleveref, autoref, thm-restate]{lipics-v2021}

\newif\ifFULL
\FULLtrue    

\usepackage{amsmath}
\usepackage{amssymb}
\usepackage{amsthm}
\usepackage{thmtools}
\usepackage{proof}
\usepackage{comment}
\usepackage{lipsum}
\usepackage{float}
\usepackage{xcolor}
\definecolor{mydeepred}{HTML}{c51b1d}
\usepackage{tikz-cd}
\usepackage{tikz}
\usetikzlibrary{arrows.meta, automata, positioning, decorations.pathmorphing, decorations.pathreplacing, shapes, calc}
\usepackage{booktabs}
\usepackage{caption}
\usepackage{pifont}
\usepackage{lscape}
\usepackage{colortbl}
\usepackage{array}
\usepackage[ruled,vlined,linesnumbered]{algorithm2e}
\usepackage{calc} 
\allowdisplaybreaks

\theoremstyle{definition}
\newtheorem{definition2}[theorem]{Definition}

\usepackage{macros}

\pdfoutput=1 
\hideLIPIcs  

\title{Efficient Matching of Some Fundamental \\ Regular Expressions with Backreferences} 

\titlerunning{Efficient Matching of Some Fundamental Regular Expressions with Backreferences} 

\author{Taisei Nogami}{Waseda University, Tokyo, Japan}{sora410@fuji.waseda.jp}{https://orcid.org/0009-0002-2820-8615}{}

\author{Tachio Terauchi}{Waseda University, Tokyo, Japan \and \url{https://www.f.waseda.jp/terauchi/}}{terauchi@waseda.jp}{https://orcid.org/0000-0001-5305-4916}{}

\authorrunning{T. Nogami and T. Terauchi}

\Copyright{Taisei Nogami and Tachio Terauchi}



\category{} 

\ifFULL
\else
\relatedversiondetails[cite={DBLP:journals/corr/abs-2504-18247}]{Full Version}{https://arxiv.org/abs/2504.18247}
\fi

\supplementdetails[cite={analysis_script}, subcategory={Analysis Script}]{Software}{https://github.com/nogamita/MFCS2025} 

\nolinenumbers 

\funding{This work was supported by JSPS KAKENHI Grant Numbers JP25KJ2137, JP23K24826, and JP20K20625.}

\EventEditors{Pawe\l{} Gawrychowski, Filip Mazowiecki, and Micha\l{} Skrzypczak}
\EventNoEds{3}
\EventLongTitle{50th International Symposium on Mathematical Foundations of Computer Science (MFCS 2025)}
\EventShortTitle{MFCS 2025}
\EventAcronym{MFCS}
\EventYear{2025}
\EventDate{August 25--29, 2025}
\EventLocation{Warsaw, Poland}
\EventLogo{}
\SeriesVolume{345}
\ArticleNo{91}

\begin{document}
\maketitle

\begin{abstract}
    Regular expression matching is of practical importance due to its widespread use in real-world applications.
    In practical use, regular expressions are often used with real-world extensions.
    Accordingly, the matching problem of regular expressions with real-world extensions has been actively studied in recent years, yielding steady progress.
    However, \emph{backreference}, a popular extension supported by most modern programming languages such as Java, Python, JavaScript and others in their standard libraries for string processing, is an exception to this positive trend.  
    In fact, it is known that the matching problem of regular expressions with backreferences (rewbs) is theoretically hard and the existence of an asymptotically fast matching algorithm for arbitrary rewbs seems unlikely. 
    Even among currently known partial solutions, the balance between efficiency and generality remains unsatisfactory.
    To bridge this gap, we present an efficient matching algorithm for rewbs of the form $e_0 (e)_1 e_1 \backslash 1 e_2$ where $e_0, e, e_1, e_2$ are pure regular expressions, which are fundamental and frequently used in practical applications.  
    It runs in quadratic time with respect to the input string length, substantially improving the best-known cubic time complexity for these rewbs.
    Our algorithm combines ideas from both stringology and automata theory in a novel way.  We leverage two techniques from automata theory, \emph{injection} and \emph{summarization}, to simultaneously examine matches whose backreferenced substrings are either a fixed \emph{right-maximal repeat} or its \emph{extendable prefixes}, which are concepts from stringology.  By further utilizing a subtle property of extendable prefixes, our algorithm correctly decides the matching problem while achieving the quadratic-time complexity.
\end{abstract}

\section{Introduction}
\label{sec:intro}

A regular expression is a convenient way to specify a language (i.e.,~a set of strings) using concatenation~($\cdot$), disjunction~($\mid$) and iteration~($\ast$).
The \emph{regular expression matching problem} (also known as \emph{regular expression membership testing}) asks whether a given string belongs to the language of a given regular expression.  
This problem is of practical importance due to its widespread use in real-world applications, particularly in format validation and pattern searching.
In 1968, Thompson presented a solution to this problem that runs in $\asympO{nm}$ time where $n$ denotes the length of the input string and $m$ the length of the input regular expression~\cite{thompson1968programming}.  We refer to his method, which constructs a nondeterministic finite automaton (NFA) and simulates it, as \emph{NFA simulation}.  

The matching problem of real-world regular expressions becomes increasingly complex due to its practical extensions such as \emph{lookarounds} and \emph{backreferences}.
Unfortunately, reducing the matching of real-world regular expressions to that of pure ones is either inefficient or impossible.
In fact, although adding positive lookaheads (a type of lookaround) does not increase the expressive power of regular expressions~\cite{morihata2012translation,berglund2021regular}, the corresponding NFAs can inevitably become enormous~\cite{miyazaki2019derivatives}.  Worse still, adding backreferences makes regular expressions strictly more expressive, meaning that an equivalent NFA may not even exist.\footnote{The rewb $( (a|b)^\ast )_1 \mybs 1$ specifies $\{ ww \mid w \in \{ a, b \}^\ast \}$, which is non-context-free (and therefore non-regular).}

Nevertheless, most modern programming languages, such as Java, Python, JavaScript and more, support lookarounds and backreferences in their standard libraries for string processing.  The most widely used implementation for the real-world regular expression matching is \emph{backtracking}~\cite{spencer1994backtracking}, an algorithm that is easy to implement and extend.  
On the other hand, the backtracking implementation suffers from a major drawback in that it takes exponential time in the worst case with respect to the input string length.  This exponential-time behavior poses the risk of \emph{ReDoS (regular expression denial of service)}, a type of DoS attack that exploits heavy regular expression matching to cause service downtime, making it a critical security issue (refer to Davis et al.~\cite{davis2018impact} for details on its history and case studies).  In response, RE2, a regular expression engine developed by Google, has deferred supporting lookarounds and backreferences, thereby ensuring $\asympO{nm}$ time complexity using NFA simulation.\footnote{The development team declares, ``Safety is RE2's raison d'\^etre.''~\cite{whyre2}}

Regarding lookarounds, several recent papers have proposed groundbreaking $\asympO{nm}$-time solutions to the matching problem of regular expressions with lookarounds~\cite{mamouras2024efficient,fujinami2024efficient,barriere2024linear}, yet regarding backreferences, the outlook is bleak.
The matching problem of regular expressions with backreferences (rewbs for short) is well known for its theoretical difficulties.
Aho showed that the rewb matching problem is NP-complete~\cite{aho1991bref}.
Moreover, rewbs can be considered a generalization of Angluin's \emph{pattern languages} (also known as \emph{patterns with variables})~\cite{angluin1979finding}; even when restricted to this, its matching problem is NP-complete with respect to the lengths of both a given string and a given pattern~\cite{angluin1979finding,ehrenfreucht1979hom,schmid2013note}, and its NP-hardness~\cite{fernau2015pattern} as well as W[1]-hardness~\cite{fernau2016parameterised} are known for certain fixed parameter settings.
The best-known matching algorithm for rewbs with at most $k$ capturing groups runs in $\asympO{n^{2k+2}m}$ time~\cite{schmid2016characterising,schmid2019regular} (With a slight modification, the time complexity can be reduced in $\asympO{n^{2k+1}m}$ time; see \cref{sec:related}).

Therefore, the existence of an efficient worst-case time complexity algorithm that works for any rewb seems unlikely, necessitating researchers to explore efficient algorithms that work for some subset of rewbs~\cite{reidenbach2011polynomial,schmid2019regular,freydenberger2019deterministic}.
Nevertheless, all existing solutions either have high worst-case time complexity or impose non-trivial constraints on the input rewbs, and finding a good balance between efficiency and generality remains an open issue.

To bridge this gap, we present an efficient matching algorithm for rewbs of the form $e_0 (e)_1 e_1 \mybs 1 e_2$, where $e_0,e,e_1,e_2$ are pure regular expressions, which are fundamental and frequently encountered in practical applications.  
While the best-known algorithm for these rewbs is the one stated above and it takes $\asympO{n^4m}$ (or $\asympO{n^3m}$) time because $k = 1$ for these rewbs, our algorithm runs in $O(n^2m^2)$ time, improving the best-known time complexity for these rewbs with respect to the input string length $n$ from cubic to quadratic.
The key appeal of this improvement lies in the replacement of the input string length $n$ with the expression length $m$.  Because $n$ is typically much larger than $m$, this improvement is considerable.

These rewbs are of both practical and theoretical interest.
From a practical perspective, these rewbs account for a large proportion of the actual usage of backreferences.  
In fact, we have confirmed that, among the dataset collected in a large-scale empirical study conducted by Davis et al.~\cite{davis2018impact}, which consists of real-world regular expressions used in npm and PyPI projects, approximately 57\% (1,659/2,909) of the non-pure rewbs\footnote{Non-pure rewbs are rewbs that use backreferences.  Note that rewbs, in general, also include ones that do not use backreferences (i.e., pure regular expressions).} are of this form~\cite{analysis_script}.

Additionally, the matching problem of rewbs of this form is a natural generalization of a well-studied foundational problem, making it theoretically interesting.
A \emph{square} (also known as \emph{tandem repeat}) is a string $\alpha \alpha$ formed by juxtaposing the same string $\alpha$. 
The problem of deciding whether a given string contains a square is of interest in stringology, and has been well studied~\cite{main1985linear,crochemore1986transducers}.
The rewb matching considered in this paper can be viewed as a generalization of the problem by regular expressions.\footnote{The problem is an instance of our rewb matching problem where $e_0,e,e_2 = \Sigma^\ast$ and  $e_1 = \varepsilon$.}

We now provide an overview of our new algorithm.  A novel aspect of the algorithm is that, \emph{unlike previous algorithms for rewb matching or square finding, it combines ideas from both stringology and automata theory}.
Our algorithm utilizes the \emph{suffix array} of the input string to efficiently enumerate candidates for backreferenced substrings (i.e.,~the contents of $\mybs 1$).
Once a candidate $\alpha$ is fixed, the matches whose backreferenced substring is $\alpha$ can be examined in $\asympO{nm}$ time in almost the same way as NFA simulation.  Because the number of candidate substrings is $\asympTheta{n^2}$, this gives an $\asympO{n^3 m}$-time algorithm (see \cref{rem:cubic} for details).
Next, we improve this baseline algorithm by extending the NFA simulation to simultaneously examine all matches whose backreferenced substrings are either a \emph{right-maximal repeat} or its \emph{extendable prefixes}, instead of examining each candidate individually.
Because the new NFA simulation requires $\asympO{m^2}$ time at each step and the number of right-maximal repeats is at most $n-1$, our algorithm runs in $\asympO{n^2m^2}$ time.

A key challenge is how to do all the examinations within time linear in $n$ for each fixed right-maximal repeat $\alpha$. To address this, we incorporate two techniques from automata theory, \emph{injection} and \emph{summarization}. Each of these techniques is fairly standard on its own (see \cref{sec:related} for their applications in prior work), but the idea of combining them is, to our knowledge, novel.
Additionally, we leverage a subtle property of $\alpha$-extendable prefixes to do the examinations correctly within time linear in $n$ even when occurrences of $\alpha$ may overlap (see \cref{subsec:ov}).

The rest of the paper is organized as follows.  \cref{sec:prelim} defines the key concepts in this paper, namely NFA simulation and right-maximal repeats.  \cref{sec:main} presents our algorithm, which is the main contribution of this paper.  \cref{sec:related} discusses related work and \cref{sec:conc} presents the conclusion and future work.
Omitted proofs are available in \fullversion{the appendix}{the full version~\cite{DBLP:journals/corr/abs-2504-18247}}.

\section{Preliminaries}
\label{sec:prelim}

Let $\Sigma$ be a set called an \emph{alphabet}, whose elements are called \emph{characters}.  
A \emph{string} $w$ is a finite sequence $a_1 \cdots a_n$ of characters $a_1, \dots, a_n$, and we write $|w|$ for the number $n$ of characters in the sequence.  The empty string is written as $\varepsilon$.
For integers $i,j \geq 0$, we write $\myint{i}{j}$ for the set of integers between $i$ and $j$.
For $i,j \in \myint{1}{n}$, we write $w[i..j]$ for the substring $a_i \cdots a_j$.  In particular, (1) $w[i] \defeq w[i..i] = a_i$ is called the \emph{character at position $i$}, (2) $w[..i] \defeq w[1..i]$ is called the \emph{prefix} up to position $i$ and (3) $w[i..] \defeq w[i..n]$ is called the \emph{suffix} from position $i$.
We regard $w[i+1..i]$ as $\varepsilon$.
A \emph{regular expression} $e$ and its language denoted by $L(e)$ are defined in the standard way.

First, we define the matching problem for rewbs of the form mentioned earlier. 
\begin{definition2}
    Given regular expressions $e_0, e, e_1, e_2$, the language of rewb $r = e_0 (e)_1 e_1 \mybs 1 e_2$, denoted by $L(r)$, is $\{ w_0 \alpha w_1 \alpha w_2 \mid w_i \in L(e_i) (i = 0, 1, 2), \alpha \in L(e) \}$.
\end{definition2}
A regular expression $e$ \emph{matches} a string $w$ if $w \in L(e)$.
The \emph{matching problem} for regular expressions is defined to be the problem of deciding whether a given regular expression matches a given string, and similarly for rewbs (of the form considered in this paper).
\begin{remark} \label{rem:rewb}
Note that in full, rewbs may use a \emph{capturing group} $(r)_i$ to assign a label $i$ to a string that the captured subexpression $r$ matches and a \emph{reference} $\mybs i$ to denote the expression that matches only the string labeled $i$.
    Therefore, rewbs are capable of more versatile expressions, such as using reference more than once (e.g.,~$(a^\ast)_1 \mybs 1 \mybs 1$) or using multiple capturing groups (e.g.,~$(a^\ast)_1 (b^\ast)_2 \mybs 1 \mybs 2$)).  For the full syntax and semantics of rewbs, refer to \cite{freydenberger2019deterministic}.  We refer to \cite{DBLP:journals/tcs/BerglundM23,DBLP:conf/mfcs/NogamiT23,DBLP:journals/iandc/NogamiT25} for studies on their expressive power.
\end{remark}

Next, we review a classical solution of the regular expression matching.
\begin{definition2}
    A \emph{nondeterministic finite automaton} (NFA) $N$ is a tuple $(Q, \delta, q_0, F)$ where $Q$ is a finite set of \emph{states}, $\delta\colon Q \times (\Sigma \cup \{\varepsilon\}) \to \mypow(Q)$ is a \emph{transition relation}, $q_0 \in Q$ is an \emph{initial state}, $F \subseteq Q$ is a set of \emph{accept states}. 
\end{definition2}
The transitive closure of a transition relation $\delta$ with the second argument fixed at $\varepsilon$ is called \emph{$\varepsilon$-closure operator} and written as $\myecl$.  Further, we lift $\myecl\colon Q \to \mypow(Q)$ to $\mypow(Q) \to \mypow(Q)$ by taking unions, i.e., $\myecl(S) \defeq \bigcup_{q\in S}\myecl(q)$.
For a state $q$ and a character $a$, we define $\Delta(q,a)$ as $\myecl(\delta(q,a))$, which informally consists of states reachable by repeating $\varepsilon$-moves from states that are reachable from $q$ by $a$.  As before, we extend $\Delta$ by setting $\Delta(S,a) \defeq \bigcup_{q\in S}\Delta(q,a)$.  Also, for a string $w$, we define $\Delta(S,w)$ as $\Delta(S,\varepsilon) \defeq S$ and $\Delta(S, wa) \defeq \Delta(\Delta(S, w), a)$.
Thus, the \emph{language} $L(N)$ of an NFA $N$ is the set of strings $w$ such that $\Delta(\myecl(q_0), w) \cap F \neq \emptyset$.

The \emph{NFA simulation} of $N$ on a string $w$ of length $n$ is the following procedure for calculating $\Delta(\myecl(q_0),w)$~\cite{thompson1968programming}.  
First, $S^{(0)} \defeq \myecl(q_0)$ is calculated, and then $S^{(i)} \defeq \Delta(S^{(i-1)}, w[i])$, called the \emph{simulation set at position $i$}, is sequentially computed from $S^{(i-1)}$ for each $i \in \myint{1}{n}$. We have $w \in L(N) \iff S^{(n)} \cap F \neq \emptyset$.

This gives a solution to the regular expression matching in $\asympO{nm}$ time and $\asympO{m}$ space where $n$ is the length of the input string $w$ and $m$ that of the input regular expression $e$. 
First, convert $e$ to an equivalent NFA $N_e$ whose number of states and transitions are both $\asympO{m}$ using the standard construction, then run the NFA simulation of $N_e$ on $w$.
Finally, check whether the last simulation set $S^{(n)}$ contains any accept state of $N_e$.
Each step of the NFA simulation can be done in $\asympO{m}$ time using breadth-first search.  Also, the procedure can be implemented in $\asympO{m}$ space by reusing the same memory for each $S^{(i)}$.

\begin{remark} \label{rem:acctest}
We call \emph{acceptance testing} the disjointness testing of a simulation set $S$ and a set $F$ of accept states, as performed above.  An acceptance test \emph{succeeds} if $S \cap F \neq \emptyset$.

Acceptance testing of an intermediate simulation set is also meaningful.
We can check whether $e$ matches each prefix $w[..i]$ with the same asymptotic complexity by testing at each position $i$.  The same can be done for the suffixes by reversing $e$ and $w$.
\end{remark}

%
Next, we review \emph{right-maximal repeats} and \emph{extendable prefixes}.  Let $w$ be a string.  A nonempty string $\alpha$ \emph{occurs} at position $i$ in $w$ if $w[i..i+|\alpha|-1] = \alpha$.  Two distinct occurrences of $\alpha$ at positions $i < j$ \emph{overlap} if $j < i + |\alpha|$.  A \emph{repeat} of $w$ is a nonempty string that occurs in $w$ more than once.  
%
A repeat $\alpha$ is called \emph{right-maximal} if $\alpha$ occurs at two distinct positions $i, j$ such that the right-adjacent characters are different (i.e., $w[i+|\alpha|] \neq w[j+|\alpha|]$).
In the special case when $\alpha$ occurs at the right end of $w$, we consider it to have a right-adjacent character distinct from all other characters when defining right-maximality.
For example, the right-maximal repeats of \mystr{mississimiss} are \mystr{i}, \mystr{iss}, \mystr{issi}, \mystr{miss}, \mystr{s}, \mystr{si}, \mystr{ss} and \mystr{ssi} (\mystr{miss}, \mystr{iss}, \mystr{ss} are due to the special treatment).\footnote{This shows that, contrary to what the word suggests, a right-maximal repeat may contain another right-maximal repeat.  In fact, the repeat \mystr{iss} and its proper substrings \mystr{i}, \mystr{s} and \mystr{ss} are all right-maximal.}
In general, the number of repeats of a string of length $n$ is $\asympTheta{n^2}$, while that of right-maximal repeats is at most $n-1$~\cite{gusfield1997book}.

For any non-right-maximal repeat $\alpha$, all the right-adjacent positions of the occurrences of $\alpha$ have the same character.  By repeatedly extending $\alpha$ with this, we will obtain the right-maximal repeat, denoted by $\myrimp{\alpha}$. This notation $\myrimp{\alpha}$ follows that of \cite{inenaga2001line,takeda2003discovering,narisawa2017efficient}.
We define $\myrimp{\alpha} \defeq \alpha$ for a right-maximal repeat $\alpha$.
A prefix $\beta$ of $\alpha$ such that $\myrimp{\beta} = \alpha$ is called \emph{$\alpha$-extendable prefix} of $\alpha$.
Note that $\alpha$ itself is $\alpha$-extendable.
Two distinct occurrences of an $\alpha$-extendable prefix $\beta$ are \emph{$\alpha$-separable} in $w$ if the two $\alpha$'s extended from the $\beta$'s have no overlap.

Finally, we mention the enumeration subroutine used in our algorithm.  Abouelhoda et al.~presented an $\asympO{n}$-time enumeration algorithm for right-maximal repeats~\cite{abouelhoda2004replacing}. 
By slightly modifying this, we obtain an $\asympO{n^2}$-time algorithm $\algenumrm$ that enumerates each right-maximal repeat $\alpha$ with the sorted array $\myarr{Idx}_\alpha$ of all starting positions of the occurrences of $\alpha$ (see \fullversion{\cref{app:enum}}{the full version~\cite{DBLP:journals/corr/abs-2504-18247}} for details).  We call $\myarr{Idx}_\alpha$ the \emph{occurrence array} of $\alpha$ in $w$.

\section{Efficient Matching}
\label{sec:main}

In this section, we show an efficient matching algorithm for rewbs of the form $e_0 (e)_1 e_1 \mybs 1 e_2$, which is the main contribution of this paper.
\begin{theorem} \label{thm:main}
    The matching problem for rewbs of the form $e_0 (e)_1 e_1 \mybs 1 e_2$, where $e_0,e,e_1,e_2$ are regular expressions, can be solved in $\asympO{n^2 m^2}$ time and $\asympO{n + m^2}$ space.
    Here, $n$ denotes the length of the input string and $m$ that of the input rewb.
    More precisely, let $\reglenleft{}, \reglencapd{}, \reglenmid{}, \reglenright{}$ denote the length of $e_0, e, e_1, e_2$ respectively, and $\mynumrmrep$ the number of right-maximal repeats of the input string, which is at most $n-1$.  Then, the problem can be solved in $\asympO{n^2 + n(\reglenleft{} + \reglenright{}) + \mynumrmrep n(\reglencapd{} + \reglenmid^2)}$ time and $\asympO{n+\max\{\reglenleft{}, \reglencapd{}, \reglenright{}, \reglenmid^2\}}$ space.  
\end{theorem}

Let $r$ be a rewb $e_0 (e)_1 e_1 \mybs 1 e_2$ and $w$ be a string.
The overview of our matching algorithm for $r$ and $w$ is as follows.
We assume without loss of generality that $e$ does not match $\varepsilon$ because we can check if $e$ matches $\varepsilon$ and $e_0 e_1 e_2$ matches $w$ in $\asympO{nm}$ time by ordinary NFA simulation.
We show an $\asympO{n(\reglencapd{} + \reglenmid^2)}$-time and $\asympO{n + \reglencapd{} + \reglenmid^2}$-space subprocedure $\algmatch(\alpha, \myarr{Idx}_\alpha)$ ($\algmatch(\alpha)$ for short) that takes a right-maximal repeat $\alpha$ and its occurrence array $\myarr{Idx}_\alpha$, and simultaneously examines all matches whose backreferenced substrings (i.e., the contents of $\mybs 1$) are $\alpha$-extendable prefixes of $\alpha$.
Before defining $\algmatch$, we state the property that characterizes its correctness:

\begin{lemma}[Correctness of $\algmatch$] \label{lem:match}
    Let $\alpha$ be a right-maximal repeat of $w$.  
    There exists a (not necessarily $\alpha$-extendable) prefix $\beta$ of $\alpha$ such that $e$ matches $\beta$ and $e_0 \beta e_1 \beta e_2$ matches $w$ if $\algmatch(\alpha)$ returns $\mytrue$.
    Conversely, $\algmatch(\alpha)$ returns $\mytrue$ if there exists an $\alpha$-extendable prefix $\beta$ of $\alpha$ such that $e$ matches $\beta$ and $e_0 \beta e_1 \beta e_2$ matches $w$.
\end{lemma}

\begin{remark} \label{rem:asymm}
Note that, interestingly, the correctness is \emph{incomplete on its own}.  That is, when there is a prefix $\beta$ of $\alpha$ such that a match with $\beta$ as the backreferenced substring exists, $\algmatch(\alpha)$ is guaranteed to return $\mytrue$ if $\beta$ is $\alpha$-extendable, but it can return $\myfalse$ if $\beta$ is not $\alpha$-extendable.
    Still, the correctness of the overall algorithm $\algmain$ described below holds because it runs $\algmatch$ on \emph{every right-maximal repeat}, and a match with a non-$\alpha$-extendable prefix $\beta$ is guaranteed to be reported by another execution of $\algmatch$ (namely, by $\algmatch(\myrimp{\beta})$).  Formally, see the proof of \cref{thm:main} described below.
\end{remark}

Given this, the overall matching algorithm $\algmain$ is constructed as follows.
It first constructs an NFA $N_{e_1}$ equivalent to the middle subexpression $e_1$ and two Boolean arrays $\myarr{Pre}$ and $\myarr{Suf}$, which are necessary for $\algmatch$.  A Boolean array $\myarr{Pre}$ (resp.~$\myarr{Suf}$) is the array which stores whether each prefix (resp.~suffix) of $w$ is matched by $e_0$ (resp.~$e_2$).
More precisely, $\myarr{Pre}$ and $\myarr{Suf}$ are the arrays such that $\myarr{Pre}[i] = \mytrue \iff w[..i] \in L(e_0)$ for $i \in \myint{0}{n}$ and $\myarr{Suf}[j] = \mytrue \iff w[j..] \in L(e_2)$ for $j \in \myint{1}{n+1}$.  Note that we can construct these arrays in $\asympO{n(\reglenleft{}+\reglenright{})}$ time and $\asympO{n + \max\{\reglenleft{},\reglenright{}\}}$ space as mentioned in \cref{rem:acctest}.
Then, it runs the $\asympO{n^2}$-time and $\asympO{n}$-space enumeration algorithm $\algenumrm$ from the final paragraph of the previous section.
Each time $\algenumrm$ outputs $\alpha$ and $\myarr{Idx}_\alpha$, $\algmatch$ is executed with these as input and using $N_{e_1}$, $\myarr{Pre}$ and $\myarr{Suf}$.
$\algmain$ returns $\mytrue$ if $\algmatch(\alpha)$ returns $\mytrue$ for some $\alpha$; otherwise, it returns $\myfalse$.

\begin{proof}[Proof of \cref{thm:main}]
    It suffices to show the correctness of $\algmain$, namely $\algmain$ returns $\mytrue$ if and only if $r$ matches $w$.
    If $w$ has a match for $r$, the (nonempty) backreferenced substring is a repeat $\beta$ of $w$ that $e$ matches.  Then, $\algmatch(\myrimp{\beta})$ returns $\mytrue$ by \cref{lem:match}.  The converse also follows from the lemma.
\end{proof}

In what follows, we present the detailed behavior of the subroutine $\algmatch$, incrementally progressing from simple cases to more complex ones.

\subsection{The Case of Nonoverlapping Right-Maximal Repeats}
\label{subsec:nonov}

Let $\alpha$ be a fixed right-maximal repeat of the input string $w$.
In this subsection, for simplicity, we consistently assume that \emph{no occurrences of $\alpha$ overlap with each other}.  We call such an $\alpha$ \emph{nonoverlapping right-maximal repeat}.
We first introduce in \cref{subsubsec:match1} a way to examine matches whose backreferenced substring is $\alpha$ itself, namely NFA simulation with auxiliary arrays and a technique called \emph{injection}.
Then, in \cref{subsubsec:match2}, we extend it to simultaneously examine all matches whose backreferenced substrings are $\alpha$-extendable prefixes of $\alpha$, instead of examining $\alpha$ individually.  There, in addition to injection, we use a technique called \emph{summarization}.\footnote{As noted in the introduction, these techniques are fairly standard on their own and often used without being given names (see \cref{sec:related} for details), but our uses of them are novel and we give them explicit names to clarify how and where they are used in our algorithm.}

\subsubsection{Only the right-maximal repeat itself}
\label{subsubsec:match1}

Let $r_\alpha$ denote the (pure) regular expression $e_0 \alpha e_1 \alpha e_2$.  
We give an $\asympO{n(m_e + m_{e_1})}$-time algorithm $\algmatchi(\alpha)$ that checks whether $r_\alpha$ matches $w$.
It runs the NFA simulation of $N_{e_1}$ using the arrays $\myarr{Pre}$, $\myarr{Suf}$ and $\myarr{Idx}_\alpha$ as oracles.
We begin by explaining the injection technique. It is grounded on the following property:
\begin{lemma} \label{lem:injecting}
    For any sets of states $S$ and $T$, and strings $u$ and $v$, we have $\Delta(S,uv) = \Delta(\Delta(S,u), v)$ and $\Delta(S,u) \cup \Delta(T,u) = \Delta(S \cup T, u)$.
\end{lemma}
This implies the following equation:
\begin{align*}
    \Delta(\myecl(q_0), uv) \cup \Delta(\myecl(q_0), v) &= \Delta(\Delta(\myecl(q_0),u),v) \cup \Delta(\myecl(q_0), v) \\
    &= \Delta(\Delta(\myecl(q_0),u) \cup \myecl(q_0), v).
\end{align*}
Therefore, we can simultaneously check whether $e$ matches \emph{either a string $uv$ or its suffix $v$} as follows:
in the NFA simulation on $uv$, when $u$ has been processed, replace the current simulation set $\Delta(\myecl(q_0), u)$ with the union of it and $\myecl(q_0)$, and then continue with the remaining simulation on $v$.
We call this replacement \emph{injection}.
More generally, for any positions $i_1 < i_2 < \cdots < i_l \le j$, we can check whether $e$ matches any of $w[i_1 .. j], \dots, w[i_l .. j]$ by testing the injected simulation set immediately after the character $w[j]$, namely $\Delta( \cdots \Delta( \Delta(\myecl(q_0), w[i_1 .. i_2-1]) \cup \myecl(q_0),  w[i_2 .. i_3-1] ) \cup \myecl(q_0) \cdots, w[i_l .. j])$.

\begin{algorithm}[t]
    \caption{$\algmatchi(\alpha)$ \label{alg:match1}}
    \DontPrintSemicolon
    \SetKwInOut{Input}{Input}\SetKwInOut{Output}{Correctness}
    %
    \Output{See \cref{lem:match1}.}
    $\iprev \leftarrow \mynil; \;\; S \leftarrow \emptyset; \;\; \ique \leftarrow \mynil; \;\; (\Delta, \myecl(q_0), F) \leftarrow N_{e_1}$\;
    \lIf{$e$ does not match $\alpha$}{
        \Return{$\myfalse$} \label{alg:match1:l:e}
    }
    \For{$\inext \in \myarr{Idx}_\alpha$}{ \label{alg:match1:l:foridx}
        \If{$\ique \neq \mynil$}{ \label{alg:match1:l:ifiqueisnotnil}
            \For{$i \leftarrow \iprev$ \KwTo $\inext - 1$}{ \label{alg:match1:l:foraltoal}
                \lIf{$S \neq \emptyset$}{
                    $S \leftarrow \Delta(S, w[i])$ \label{alg:match1:l:nfasim}
                }
                \lIf{$i = \ique$}{
                    $S \leftarrow S \cup \myecl(q_0)$ \tcc*[f]{Injection} \label{alg:match1:l:inj}
                }
            }
            \lIf{$S \cap F \neq \emptyset$ and $\myarr{Suf}[\inext+|\alpha|]$}{ \label{alg:match1:l:test}
                \Return{$\mytrue$} \label{alg:match1:l:rett}
            }
        }
        $\iprev \leftarrow \inext$\; \label{alg:match1:l:update}
        \lIf{$\myarr{Pre}[\iprev-1]$}{ 
            $\ique \leftarrow \iprev + |\alpha| - 1$ \label{alg:match1:l:que}
        }
    }
    \Return{$\myfalse$}
\end{algorithm}

We now explain $\algmatchi(\alpha)$ whose pseudocode is shown in \cref{alg:match1}.  First, it checks if $e$ matches $\alpha$ and returns $\myfalse$ if it is false; otherwise, the algorithm continues running.
Let $i_1 < i_2 < \cdots$ be the positions in $\myarr{Idx}_\alpha$.
Then, it searches for the starting position $i_{j_1}$ of the leftmost occurrence of $\alpha$ such that $e_0$ matches the prefix $w[..i_{j_1}-1]$ to the left of the $\alpha$ by looking at $\myarr{Pre}[i_j-1]$ sequentially for each $i_j \in \myarr{Idx}_\alpha$ (lines~\ref{alg:match1:l:foridx}, \ref{alg:match1:l:update} and \ref{alg:match1:l:que}).
Remark that the $\alpha$ at position $i_{j_1}$ is the leftmost candidate that the left $\alpha$ of $r_\alpha$ may correspond to in a match.

If $i_{j_1}$ is found, it starts an NFA simulation of $N_{e_1}$ from the position $i_{j_1} + |\alpha|$ immediately to the right of the $\alpha$ at position $i_{j_1}$ by setting $S$ to be $\myecl(q_0)$ immediately after the character $w[i_{j_1}+|\alpha|-1]$ (line~\ref{alg:match1:l:inj}).  Note that $S$ is $\emptyset$ prior to the assignment.
In what follows, let $i_{j_2}, i_{j_3}, \dots$ denote the starting positions of the $\alpha$'s to the right of the $\alpha$ at position $i_{j_1}$ in sequence.  Note that, unlike in the case of $i_{j_1}$, we do not assume that $e_0$ matches the prefix $w[..i_{j_{k-1}}]$ for $i_{j_k}$, i.e., $j_k = j_1 + k - 1$ ($k \ge 2$).

Next, the algorithm resumes the simulation and proceeds until the character $w[i_{j_2}-1]$ (lines~\ref{alg:match1:l:foraltoal} and \ref{alg:match1:l:nfasim}).
Then, it performs the acceptance testing $S \cap F \neq \emptyset$ (line~\ref{alg:match1:l:test}).  If it succeeds, $e_0 \alpha e_1 \alpha$ matches $w[..i_{j_2}+|\alpha|-1]$ by matching the two $\alpha$'s to $w[i_{j_1} .. i_{j_1} + |\alpha| - 1]$ and $w[i_{j_2} .. i_{j_2} + |\alpha| - 1]$.  Accordingly, the algorithm further checks whether $e_2$ matches the remaining suffix $w[i_{j_2}+|\alpha| ..]$ by looking at $\myarr{Suf}[i_{j_2} + |\alpha|]$.  If it is $\mytrue$, $r_\alpha$ matches $w$ and the algorithm returns $\mytrue$.  Otherwise, there is no possibility that $r_\alpha$ matches $w$ in a way that the right $\alpha$ of $r_\alpha$ matches the $\alpha$ at position $i_{j_2}$, and the algorithm continues running.

In this way, the algorithm proceeds from position $i_{j_{k-1}}$ to position $i_{j_k}$ for $k = 2,3,\dots$ as follows, while assigning $i_{j_{k-1}}$ and $i_{j_k}$ to variables $\iprev$ and $\inext$ respectively at each $k$:
(i) it processes the substring $w[\iprev .. \iprev+|\alpha|-1]$ (lines~\ref{alg:match1:l:foraltoal} and \ref{alg:match1:l:nfasim}), and then (ii) injects $\myecl(q_0)$ into the simulation set $S$ if the $\alpha$ at position $\iprev$ is a candidate that the left $\alpha$ of $r_\alpha$ may correspond to in a match (i.e., if $e_0$ matches $w[..\iprev-1]$) (line~\ref{alg:match1:l:inj}).  Next, (iii) it resumes the simulation and proceeds until the character $w[\inext-1]$ (lines~\ref{alg:match1:l:foraltoal} and \ref{alg:match1:l:nfasim}), and then (iv) performs the acceptance testing and checks whether $e_2$ matches the remaining suffix $w[\inext + |\alpha| ..]$ (line~\ref{alg:match1:l:test}).
Finally, (v) it updates $\iprev$ using $\inext$ and if the $\alpha$ at $i_{j_k}$ is a candidate that the left $\alpha$ of $r_\alpha$ may correspond to in a match, then keep its right-adjacent position $i_{j_k}+|\alpha|-1$ in $\ique$ for future injection (lines~\ref{alg:match1:l:update} and \ref{alg:match1:l:que}).
If step (iv) succeeds for some $j$, it returns $\mytrue$; otherwise, it returns $\myfalse$.

$\algmatchi{}$ runs in $\asympO{n(m_e + m_{e_1})}$ time because it runs an NFA simulation of $e$ and an NFA simulation of $e_1$ with injection.
For correctness, the following is essential.

\begin{proposition} \label{prop:match1simset}
    Let $i_1 < i_2 < \cdots$ be the positions in $\myarr{Idx}_\alpha$.
    Every time line~\ref{alg:match1:l:test} is reached at an iteration with $\inext = i_j$, we have $S = \bigcup_{j' \in J'} \Delta(\myecl(q_0), w[i_{j'}+|\alpha|, i_{j}-1])$ where $J' = \{ j' \in \myint{1}{|\myarr{Idx}_\alpha|} \mid i_{j'} + |\alpha| \le i_j \text{ and } w[..i_{j'}-1] \in L(e_0) \}$.
\end{proposition}

\begin{lemma}[Correctness of $\algmatchi$] \label{lem:match1}
    Let $\alpha$ be a nonoverlapping right-maximal repeat of $w$.  Then,
    $\algmatchi(\alpha)$ returns $\mytrue$ if and only if $e$ matches $\alpha$ and $r_\alpha$ matches $w$.
\end{lemma}

\begin{remark} \label{rem:cubic}
    In fact, $\algmatchi{}$ works correctly for any repeat $\alpha$ and not only right-maximal ones.
This gives an $O(n^3m)$-time matching algorithm for rewbs of our form by modifying $\algenumrm$ to output not only all right-maximal repeats but all repeats.
As mentioned in the introduction, we note that this time complexity itself can also be achieved by existing algorithms.
    Further improvements in time complexity require additional ideas that we describe in the following sections as extensions of the baseline algorithm $\algmatchi$.
\end{remark}

\begin{remark} \label{rem:which}
    An essential and interesting property of the algorithm is that if it returns $\mytrue$, the existence of a match is guaranteed, \emph{but we do not know where $(e)_1$ and $\mybs 1$ match}.  This is because injecting $\myecl(q_0)$ into the simulation set in an NFA simulation means identifying the current position with the starting position of the NFA simulation.
\end{remark}

\subsubsection{The extendable prefixes of the right-maximal repeat}
\label{subsubsec:match2}

Recall that $\alpha$ is a fixed nonoverlapping right-maximal repeat of $w$.
We extend $\algmatchi{}$ to simultaneously examine all matches whose backreferenced substrings are $\alpha$-extendable prefixes of $\alpha$.
To this end, we split the NFA simulation of $\algmatchi$ in two phases.

We first introduce a technique called \emph{summarization}.  Let $q_1, \dots, q_{\nfastmid{}}$ be the states of $N_{e_1}$, the NFA equivalent to the middle subexpression $e_1$ of the input rewb that we fixed earlier.\footnote{Note that $\reglenmid{}$ denotes the length of $e_1$, whereas $\nfastmid{}$ denotes the number of states in $N_{e_1}$.  However, they may be regarded as the same because they differ only by a constant factor.}
While ordinary NFA simulation starts only from the initial state, \emph{NFA simulation with summarization} (NFASS) starts its simulation from each state $q_1, \dots, q_{\nfastmid{}}$.
Consequently, the ``simulation set'' of an NFASS is a vector of simulation sets $\mathcal{S} = \langle \mathcal{S}[1], \dots, \mathcal{S}[\nfastmid{}] \rangle$, where each $\mathcal{S}[l]$ is the simulation set of an ordinary NFA simulation but with $q_l$ regarded as its initial state.
We write $\myDeltasum(\mathcal{S}, u)$ for $\langle \Delta(\mathcal{S}[1], u), \dots, \Delta(\mathcal{S}[\nfastmid{}], u) \rangle$.
Note that each step of an NFASS takes $\asympO{\nfastmid^2} = \asympO{\reglenmid^2}$ time.

\begin{algorithm}[t]
    \caption{$\algmatchii(\alpha)$ \label{alg:match2}}
    \DontPrintSemicolon
    \SetKwInOut{Input}{Input}\SetKwInOut{Output}{Correctness}
    %
    \Output{See \cref{lem:match2}.}
    $\iprev \leftarrow \mynil; \;\; \mathcal{S} \leftarrow \langle\emptyset, \dots, \emptyset\rangle; \;\; \ique \leftarrow \mynil; \;\; (Q=\{q_1,\dots,q_{\nfastmid{}}\},\myDeltasum,F) \leftarrow N_{e_1}$\; 
    Construct an array $\myarr{Pre}_\alpha$ such that $\myarr{Pre}_\alpha[k] = \mytrue \iff \alpha[..k] \in L(e)$ for $k \in \myint{1}{|\alpha|}$, which is used later in $\algintmed$\; \label{alg:match2:prealpha}
    \For{$\inext \in \myarr{Idx}_\alpha$}{ \label{alg:match2:mainfor}
        \If{$\ique \neq \mynil$}{
            \For{$i \leftarrow \iprev$ \KwTo $\inext - 1$}{
                \lIf{$\mathcal{S} \neq \langle\emptyset, \dots, \emptyset\rangle$}{
                    $\mathcal{S} \leftarrow \myDeltasum(\mathcal{S}, w[i])$ \tcc*[f]{Summarization} \label{alg:match2:nfass}
                }
                \lIf{$i = \ique$}{
                    $\mathcal{S} \leftarrow \langle \mathcal{S}[1] \cup \{q_1\}, \dots, \mathcal{S}[\nfastmid{}] \cup \{q_{\nfastmid{}}\} \rangle$ \tcc*[f]{Injection} \label{alg:match2:inj}
                }
            }
            $T \leftarrow \algintmed(\inext, \inext + |\alpha| - 1)$\;  \label{alg:match2:intmed}
            \lIf{$\exists q_l \in T. \mathcal{S}[l] \cap F \neq \emptyset$}{ \label{alg:match2:test}
                \Return{$\mytrue$}
            }
        }
        $\iprev \leftarrow \inext$\;
        \lIf{$\myarr{Pre}[\iprev-1]$}{
            $\ique \leftarrow \iprev + |\alpha| - 1$
        }
    } \label{alg:match2:mainforend}
    \Return{$\myfalse$}\;
\end{algorithm}

Building on the above, we describe $\algmatchii(\alpha)$ whose pseudocode is shown in \cref{alg:match2}.
Although the overall flow is similar to $\algmatchi$ in \cref{alg:match1}, it has two major differences.

One difference is that the NFA simulation using a simulation set $S$ has been replaced by the NFASS using $\mathcal{S}$ (line~\ref{alg:match2:nfass}).
Similarly to step (ii) of $\algmatchi$ (\cref{subsubsec:match1}), when the NFASS reaches an occurrence of $\alpha$ at position $i$ where $e_0$ matches the prefix $w[.. i-1]$, it injects $\{q_l\}$ into $\mathcal{S}[l]$ for each $l \in \myint{1}{\nfastmid{}}$ immediately after the character $w[i+|\alpha|-1]$.
Analogously to \cref{prop:match1simset}, the following proposition holds. 

\begin{proposition} \label{prop:match2simset}
    Let $i_1 < i_2 < \cdots$ be the positions in $\myarr{Idx}_\alpha$.
    Every time line~\ref{alg:match2:test} is reached at an iteration with $\inext = i_j$, we have $\mathcal{S}[l] = \bigcup_{j' \in J'} \Delta(\{q_l\}, w[i_{j'}+|\alpha|, i_{j}-1])$ where $J' = \{ j' \in \myint{1}{|\myarr{Idx}_\alpha|} \mid i_{j'} + |\alpha| \le i_j \text{ and } w[..i_{j'}-1] \in L(e_0) \}$.
\end{proposition}

\begin{algorithm}[t]
    \caption{$\algintmed(\ibeg,\iend)$ \label{alg:intmed}}
    \DontPrintSemicolon
    \SetKwInOut{Input}{Input}\SetKwInOut{Output}{Correctness}
    %
    \Output{See \cref{lem:intmed}.}
    $T \leftarrow \emptyset; \;\; (\Delta,\myecl(q_0),F) \leftarrow N_{e_1}$\;
    \For{$i \leftarrow \ibeg$ \KwTo $\iend$}{ \label{alg:intmed:for}
        \lIf{$\myarr{Pre}_\alpha[i - (\iend - |\alpha|)]$ and $\myarr{Suf}[i+1]$}{
            $T \leftarrow T \cup \myecl(q_0)$ \tcc*[f]{Injection} \label{alg:intmed:inj}
        }
        \lIf{$T \neq \emptyset$ and $i < \iend$}{
            $T \leftarrow \Delta(T, w[i+1])$ \label{alg:intmed:nfasim}
        }
    }
    \Return{$T$}\;
\end{algorithm}

The other major difference is the guard condition for the algorithm returning $\mytrue$ (line~\ref{alg:match2:test}). 
Each time the NFASS reaches the occurrence of $\alpha$ at position $i_j$, it runs another NFA simulation on the $\alpha$ with injection (line~\ref{alg:match2:intmed}).
The aim of this subsimulation is, roughly, to calculate a set of reachable states $T$ from the initial state $q_0$ of $N_{e_1}$ by any suffix $\alpha[k+1..]$ of $\alpha$ whose corresponding prefix $\alpha[..k]$ is a candidate at position $i_j$ for the content of $\mybs 1$ in a match, namely $\alpha[k+1..]$ where $e$ matches $\alpha[..k]$ and $e_2$ matches $w[i_j+k..]$.
Then, $\algmatchii(\alpha)$ composes $T$ and $\mathcal{S}$, that is, checks if there exists a state $q_l$ of $T$ such that the acceptance testing of $\mathcal{S}[l]$ succeeds (line~\ref{alg:match2:test}).
If such $q_l$ exists, we can build a match by concatenating a suffix $\alpha[k+1..]$ that takes $q_0$ to $q_l$ and a substring $u$ of $w$ that lies in two occurrences of $\alpha$ that takes $q_l$ to an accept state in $N_{e_1}$.
%
In this scenario, $r$ matches $w$ and the algorithm returns $\mytrue$; otherwise, it continues running.

\cref{alg:intmed} shows $\algintmed(\ibeg,\iend)$, the algorithm which calculates $T$ at line~\ref{alg:match2:intmed} of \cref{alg:match2}.
It uses a Boolean array $\myarr{Pre}_\alpha$ such that $\myarr{Pre}_\alpha[k] = \mytrue \iff \alpha[..k] \in L(e)$ for $k \in \myint{1}{|\alpha|}$ which is precomputed at the beginning of $\algmatchii(\alpha)$ (line~\ref{alg:match2:prealpha} of \cref{alg:match2}).

$\algintmed$ requires that $\iend$ is the right end position of an occurrence of $\alpha$ and $\ibeg$ is a position that belongs to the $\alpha$.  Note that the algorithm is presented in a generalized form to be used later in \cref{subsec:ov}, but $\algmatchii$ always passes $\inext$ to the argument $\ibeg$.
Each time it reaches a position $i \in \myint{\ibeg}{\iend}$, it checks whether $e$ matches the prefix of $\alpha$ which ends at $i$ and $e_2$ matches the remaining suffix of $w$ by looking at $\myarr{Pre}_\alpha[i - (\iend - |\alpha|)]$ and $\myarr{Suf}[i+1]$ (line~\ref{alg:intmed:inj}).
If both checks succeed, it starts an NFA simulation or injects $\myecl(q_0)$ into the ongoing simulation set $T$.
The correctness of the algorithm is as follows:

\begin{lemma}[Correctness of $\algintmed$] \label{lem:intmed}
    Let $\ibeg$ and $\iend$ be positions of $w$ where $\iend$ is the right end of some occurrence of $\alpha$ and $\iend - |\alpha| < \ibeg \leq \iend$.  Then, $\algintmed(\ibeg,\iend)$ returns $T = \bigcup_{i \in I} \Delta(\myecl(q_0),w[i+1 .. \iend])$ where $I = \{ i \in \myint{\ibeg}{\iend} \mid w[\iend - |\alpha| + 1 .. i] \in L(e) \text{ and } w[i+1 ..] \in L(e_2)\}$.
\end{lemma}


Given this, we prove the correctness of $\algmatchii$ in the following lemma.

\begin{lemma}[Correctness of $\algmatchii$] \label{lem:match2}
    Let $\alpha$ be a nonoverlapping right-maximal repeat of $w$.
    Then, there exists a prefix $\beta$ of $\alpha$ such that $e$ matches $\beta$ and $e_0 \beta e_1 \beta e_2$ matches $w$ if $\algmatchii(\alpha)$ returns $\mytrue$.
    Conversely, $\algmatchii(\alpha)$ returns $\mytrue$ if there exists an $\alpha$-extendable prefix $\beta$ of $\alpha$ such that $e$ matches $\beta$ and $e_0 \beta e_1 \beta e_2$ matches $w$.
\end{lemma}

Regarding the time complexity, $\algmatchii$ runs in $\asympO{n (m_e + m_{e_1}^2)}$ time because its main loop processes each position of $w$  at most twice (once by the NFASS and once by $\algintmed$ at line~\ref{alg:match2:intmed}) and each step takes $\asympO{m_{e_1}^2}$ time.  Note that $\algintmed$ does not revisit any position because we assumed that $\alpha$ is a nonoverlapping right-maximal repeat.

\begin{example} \label{ex:match2}
    We illustrate $\algmatchii$ with its execution on the instance defined as follows.
    Let $w = \mystr{abbabbabbabba}$ be the input string over $\Sigma = \{\mystr{a}, \mystr{b}\}$ and $e_0 (e)_1 e_1 \mybs 1 e_2$ be the input rewb, where $e = \Sigma^*$, $e_2 = (\Sigma\Sigma)^*$, $e_0$ matches the strings that contain at most two $\mystr{b}$'s, and $e_1$ matches those that contain $\mystr{b}$ at least three and an odd number of times.
    We consider the case where $\alpha = \mystr{bba}$, which is a nonoverlapping right-maximal repeat of $w$.
    Note that its occurrence array $\myarr{Idx}_\alpha$ is $[2,5,8,11]$.

    \begin{figure}[t]
        \centering
        \includegraphics[scale=0.52]{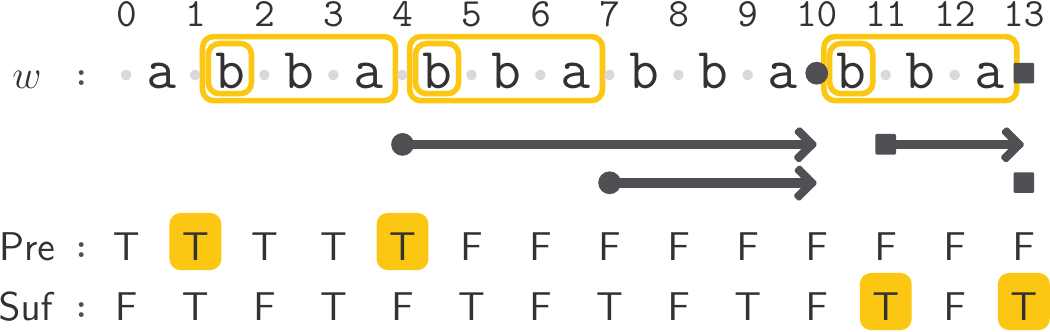}
        \caption{An example execution of $\algmatchii$.}
        \label{fig:match2}
    \end{figure}

    \cref{fig:match2} shows the point at which $\algintmed$, which was called at line~\ref{alg:match2:intmed}, has completed its execution during the loop for $\inext = 11$.  The top row shows the position of $w$, and the two rows labeled $\myarr{Pre}$ and $\myarr{Suf}$ represent the Boolean values of the corresponding arrays at each position.
    Within the row of $w$, the bullet and the square mark the positions being currently processed by the NFASS and the execution of $\algintmed$, respectively.
    Analogously, the arrows below $w$, starting from the bullets and the squares, indicate the past behaviors of those.  
    
    The bullet at position~4 marks where the NFASS started because the $\alpha$ at position~2 is the leftmost among the occurrences of $\alpha$ that $e_0$ matches the prefix of $w$ to the left (i.e., $\myarr{Pre}[1] = \mytrue$).
    Similarly, the one at position~7 marks where injection was performed in the NFASS because the $\alpha$ at position~5 was such an $\alpha$ (i.e., $\myarr{Pre}[4] = \mytrue$).
    Also, the squares at positions~11 and 13 mark where the execution of $\algintmed$ started and injection was performed in it because $\mystr{b}$ and $\mystr{bba}$ are prefixes of $\alpha$ that $e$ matches (which always holds in this instance) and $e_2$ matches whose remaining suffix of $w$ (i.e., $\myarr{Suf}[11] = \myarr{Suf}[13] = \mytrue$), respectively.
    
    Then, the algorithm checks at line~\ref{alg:match2:test} if $e_1$ matches any of $w[3..10]$, $w[5,10]$, $w[6..10]$ and $w[8,10]$, and returns $\mytrue$ because $e_1$ matches $w[3..10]$ and $w[6..10]$.  Observe that, as stated in \cref{rem:which}, it cannot determine which of the four $e_1$ actually matches.
\end{example}

\subsection{The General Case of Right-Maximal Repeats}
\label{subsec:ov}

Let $\alpha$ be a right-maximal repeat.
In this subsection, we give the full version of our algorithm $\algmatch(\alpha)$ that works for the general case where $\alpha$ is possibly overlapping.

We first explain why the aforementioned algorithm $\algmatchii(\alpha)$ does not work correctly in this case.
%
There are two main reasons.
One is the time complexity.  Note that in $\algmatchii(\alpha)$, $\algintmed$ is called at every position right before where $\alpha$ occurs (line~\ref{alg:match2:intmed} of \cref{alg:match2}).  Under the nonoverlapping assumption, the total time $\algintmed$ takes is linear in $n$ because the total length of all occurrences of $\alpha$ is also linear, but that does not necessarily hold when $\alpha$ may overlap.
The other reason concerns the correctness of the algorithm. 
When overlaps are allowed, there may exist a match whose backreferenced substring occurs as nonoverlapping $\alpha$-extendable prefixes of some overlapping occurrences of $\alpha$.
Because $\algmatchii$ is not designed to find such matches, it may falsely report that no match exists.

The key observation to overcome these obstacles is, as stated in \cref{rem:asymm}, that it only needs to check if there are matches whose backreferenced substrings are $\alpha$-extendable prefixes of $\alpha$, rather than arbitrary prefixes of $\alpha$.
The following lemma gives a necessary condition for a prefix of $\alpha$ being $\alpha$-extendable (a related statement appears as Lemma~5 in \cite{takeda2003discovering}):
\begin{lemma} \label{lem:once}
    Let $\alpha$ be a right-maximal repeat.
    Suppose that $\alpha$ contains its prefix $\beta$ at least twice: once as a prefix and once elsewhere. Then $\beta$ is a non-$\alpha$-extendable prefix of $\alpha$.
    More generally, if two occurrences of $\alpha$ have an overlap of length $d$, the prefixes of $\alpha$ whose length is no more than $d$ are non-$\alpha$-extendable prefixes of $\alpha$.
\end{lemma}

In what follows, we divide the algorithm $\algmatch$ into two subalgorithms $\algmatchiiia$ and $\algmatchiiib$, and describe how they address the obstacles noted above.
$\algmatchiiia(\alpha)$ (resp.~$\algmatchiiib(\alpha)$) is an algorithm to detect a match whose backreferenced substring occurs as an $\alpha$-extendable prefix of some nonoverlapping occurrences (resp.~a nonoverlapping $\alpha$-extendable prefix of some overlapping occurrences) of $\alpha$.
Consequently, $\algmatch(\alpha)$ is an algorithm that returns $\mytrue$ if and only if at least one of the subalgorithms returns $\mytrue$.

\begin{algorithm}[t]
    \caption{$\algmatchiiia(\alpha)$ \label{alg:match3a}}
    \DontPrintSemicolon
    \SetKwInOut{Input}{Input}\SetKwInOut{Output}{Correctness}
    \Output{See \cref{lem:match3a}.}
    %
    %
    $\iprev \leftarrow \mynil; \;\; \mathcal{S} \leftarrow \langle\emptyset, \dots, \emptyset\rangle; \;\; \myarr{Que} \leftarrow \mynil; \;\; (Q=\{q_1,\dots,q_{\nfastmid{}}\},\myDeltasum,F) \leftarrow N_{e_1}$\;
    Construct an array $\myarr{Pre}_\alpha$ such that $\myarr{Pre}_\alpha[k] = \mytrue \iff \alpha[..k] \in L(e)$ for $k \in \myint{1}{|\alpha|}$, which is used later in $\algintmed$\;
    $d \leftarrow \max (\{0\}\cup \{ \myarr{Idx}_\alpha[j-1] + |\alpha| - \myarr{Idx}_\alpha[j] \mid j \ge 2 \})$\; \label{alg:match3a:d}
    \For{$\inext \in \myarr{Idx}_\alpha$}{
        \If{$\myarr{Que} \neq \mynil$}{ \label{alg:match3a:qg}
            \For{$i = \iprev$ \KwTo $\inext - 1$}{ \label{alg:match3a:iptoin}
                \lIf{$\mathcal{S} \neq \langle \emptyset, \dots, \emptyset \rangle$}{
                    $\mathcal{S} \leftarrow \myDeltasum(\mathcal{S}, w[i])$ \tcc*[f]{Summarization}
                }
                \If{$\myarr{Que} \neq []$ and $\myarr{Que}.top = i$}{
                    $\mathcal{S} \leftarrow \langle \mathcal{S}[1] \cup \{q_1\}, \dots, \mathcal{S}[\nfastmid{}] \cup \{q_{\nfastmid{}}\} \rangle$ \tcc*{Injection}
                    $\myarr{Que}.dequeue()$\; \label{alg:match3a:dequeue}
                }
            }
            $T \leftarrow \algintmed(\inext{} + d, \inext+|\alpha|-1)$\;  \label{alg:match3a:intmed}
            \lIf{$\exists q_l \in T. \mathcal{S}[l] \cap F \neq \emptyset$}{ 
                \Return{$\mytrue$} \label{alg:match3a:test}
            }
        }
        $\iprev \leftarrow \inext$\;
        \If{$\myarr{Pre}[\iprev-1]$}{ \label{alg:match3a:preiprev}
            \lIf{$\myarr{Que} = \mynil$}{$\myarr{Que} \leftarrow []$}
            $\myarr{Que}.enqueue(\iprev + |\alpha| - 1)$\; \label{alg:match3a:enqueue}
        }
    }
    \Return{$\myfalse$}\;
\end{algorithm}

\cref{alg:match3a} shows $\algmatchiiia(\alpha)$, the subalgorithm for detecting a match whose backreferenced substring occurs as an $\alpha$-extendable prefix of some nonoverlapping occurrences of $\alpha$.
We explain the changes from $\algmatchii$.
Recall the first obstacle: $\algintmed$ takes too much time.
Let $d$ be the maximum length of the overlapping substrings of the occurrences of $\alpha$, i.e., $d = \max(\{ 0 \} \cup\{ \myarr{Idx}_\alpha[j-1] + |\alpha| - \myarr{Idx}_\alpha[j] \mid j \ge 2 \})$.\footnote{$\algenumrm$ always returns $\myarr{Idx}_\alpha$ of size at least 2.  Note that an $\alpha$ that occurs less than twice need not be considered.} $\algmatchiiia$ precomputes $d$ at line~\ref{alg:match3a:d}.  Clearly, this can be done in $\asympO{n}$ time by only considering each two adjacent occurrences of $\alpha$.
Then, by \cref{lem:once}, the algorithm only needs to examine a match whose backreferenced substring is a prefix of $\alpha$ strictly longer than $\alpha[..d]$, namely any of $\alpha[..d+1], \dots, \alpha[..|\alpha|] = \alpha$.
Therefore, $\algmatchiiia$ calls $\algintmed$ so that it starts from position $\inext + d$ rather than $\inext$ (line~\ref{alg:match3a:intmed}), ensuring its $\asympO{n(m_{e} + m_{e_1}^2)}$ time complexity.

A subtle issue here is that in the NFASS of $\algmatchiiia$, there may be multiple timings of injection before reaching another occurrence of $\alpha$.
This is dealt with by managing them with an FIFO structure $\myarr{Que}$ instead of a variable $\ique$ as was done in $\algmatchii$.

The following lemma states the correctness of $\algmatchiiia$.
Recall the definition of $\alpha$-separability from \cref{sec:prelim}.

\begin{lemma}[Correctness of $\algmatchiiia$] \label{lem:match3a}
    Let $\alpha$ be a right-maximal repeat of $w$.
    There exists a prefix $\beta$ of $\alpha$ such that $e$ matches $\beta$ and $e_0 \beta e_1 \beta e_2$ matches $w$ if $\algmatchiiia(\alpha)$ returns $\mytrue$.
    Conversely, $\algmatchiiia(\alpha)$ returns $\mytrue$ if there exists an $\alpha$-extendable prefix $\beta$ of $\alpha$ such that (i) $e$ matches $\beta$, (ii) $e_0 \beta e_1 \beta e_2$ matches $w$ and (iii) the two occurrences of $\beta$ are $\alpha$-separable in $w$.
\end{lemma}

\begin{figure}[t]
    \begin{subfigure}{0.5\linewidth}
        \centering
        \includegraphics[scale=0.39]{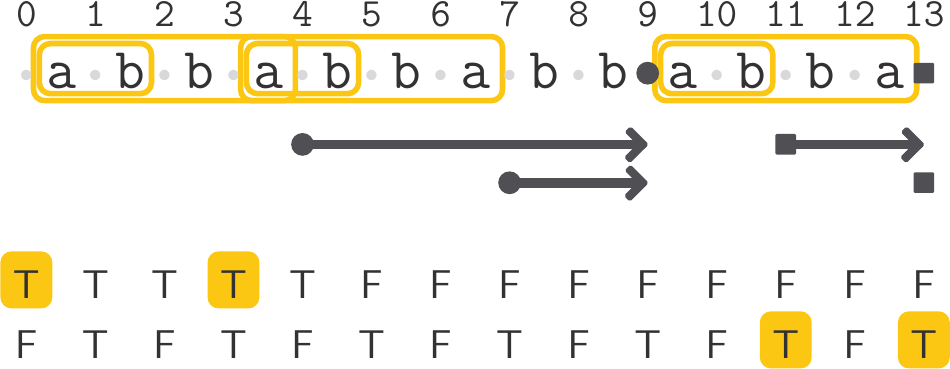}
        \caption{\centering $\algmatchiiia$.}
        \label{subfig:match3a}
    \end{subfigure}\hfill
    \begin{subfigure}{0.5\linewidth}
        \centering
        \includegraphics[scale=0.39]{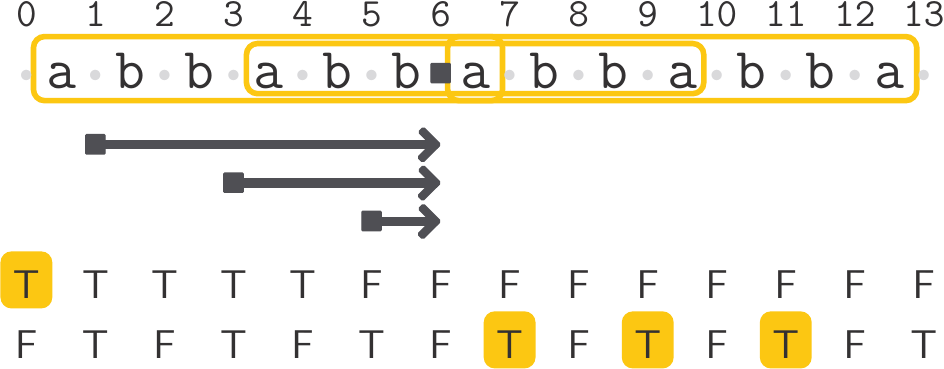}
        \caption{\centering $\algmatchiiib$.}
        \label{subfig:match3b}
    \end{subfigure}
    \caption{Example executions. The meaning of the rows is the same as in \cref{fig:match2}.}
    \label{fig:match3}
\end{figure}

\begin{example} \label{ex:match3a}
      \cref{fig:match3}(\subref{subfig:match3a}) shows the execution of the algorithm on the same instance as \cref{ex:match2} where $\alpha = \mystr{abba}$, which is an overlapping right-maximal repeat of $w$ and whose occurrence array $\myarr{Idx}_\alpha$ is $[1,4,7,10]$.  
It returns $\mytrue$ because $e_1$ matches $w[3..9]$ and $w[6..9]$.
Observe that it skips the check for a match whose backreferenced substring is $\mystr{a}$ because $\mystr{a}$ is non-$\alpha$-extendable by $d = 1$ and \cref{lem:once}.  In fact, $\mystr{a}$ is right-maximal and the check is instead performed by the execution of $\algmatchiiia$ with $\alpha = \mystr{a}$, as mentioned in \cref{rem:asymm}.
\end{example}

\begin{algorithm}[t]
    \caption{$\algmatchiiib(\alpha)$ \label{alg:match3b}}
    \DontPrintSemicolon
    \SetKwInOut{Input}{Input}\SetKwInOut{Output}{Correctness}
    \SetKwProg{Proc}{procedure}{}{end procedure}
    \SetKwFunction{FnZip}{zip}
    \Output{See \cref{lem:match3b}.}
    \tcc{zip($\myarr{A},\myarr{B}$) = $[\langle \myarr{A}[j], \myarr{B}[j] \rangle \mid 1 \leq j \leq |\myarr{A}| ]$ provided that $|\myarr{A}| = |\myarr{B}|$}
    %
    %
    $\myarr{Fwd} \leftarrow []; \;\; \jleft \leftarrow |\myarr{Idx}_\alpha|; \;\; \jright \leftarrow |\myarr{Idx}_\alpha|$\; \label{alg:match3b:belbegin}
    \While{$\jleft \ge 1$}{
        \If{$\myarr{Idx}_\alpha[\jright] \le \myarr{Idx}_\alpha[\jleft] + |\alpha| - 1$}{
            $\myarr{Fwd}[\jleft] \leftarrow \myarr{Idx}_\alpha[\jright]$\; \label{alg:match3b:belasgn}
            $\jleft \leftarrow \jleft - 1$\;
        }
        \lElse{
            $\jright \leftarrow \jright - 1$
        }
    } \label{alg:match3b:belend}
    $\fprev \leftarrow 0; \;\; S \leftarrow \emptyset; \;\; (\Delta,\myecl(q_0),F) \leftarrow N_{e_1}$\;
    \For{$\langle \inext, \fnext \rangle \in \FnZip{$\myarr{Idx}_\alpha, \myarr{Fwd}$}$}{  \label{alg:match3b:mainfor}
        \If{$\inext < \fnext$, $\myarr{Pre}[\inext-1]$ and $\fprev < \fnext$}{ \label{alg:match3b:valleft}
            $S \leftarrow \emptyset$\; \label{alg:match3b:reset}
            \For{$i \leftarrow \max\{\inext,\fprev\}$ \KwTo $\fnext-1$}{ \label{alg:match3b:for}
                \If{$\myarr{Pre}_\alpha[i - \inext + 1]$ and $\myarr{Suf}[\fnext+i-\inext+1]$}{
                    $S \leftarrow S \cup \myecl(q_0)$ \tcc*{Injection}
                }
                \lIf{$S \neq \emptyset$ and $i < \fnext - 1$}{
                    $S \leftarrow \Delta(S, w[i+1])$ \label{alg:match3b:nfasim}
                }
            }
            \lIf{$S \cap F \neq \emptyset$}{
                \Return{$\mytrue$} \label{alg:match3b:test}
            }
        }
        $\fprev \leftarrow \fnext$\; \label{alg:match3b:fupdate}
    }
    \Return{$\myfalse$}\;
\end{algorithm}

Next, we explain $\algmatchiiib(\alpha)$ shown in \cref{alg:match3b}, the subalgorithm for detecting a match whose backreferenced substring occurs as a nonoverlapping $\alpha$-extendable prefix of some overlapping occurrences of $\alpha$.

We first explain the challenges with detecting such matches in time linear in $n$.
Fix an occurrence of $\alpha$ and suppose that no other $\alpha$ overlaps it from the left and it overlaps other $\alpha$'s to the right.
We name them $\alpha_1, \alpha_2, \alpha_3, \dots, \alpha_{k}$ from left to right with $\alpha_1$ being the one we fixed earlier.
It may seem that $\algmatchiiib(\alpha)$ has to check matches between every pair of these $\alpha_i$'s.  Doing this naively takes $\asympTheta{k^3}$ time, and as $k = \asympTheta{n}$ in general, this becomes $\asympTheta{n^3}$ time (for example, the case $w = \mystr{a}^{2n}$ and $\alpha_i = w[i..i+n-1]$ for $i \in \myint{1}{n}$).

Our key observation is that $\algmatchiiib(\alpha)$ actually only needs to examine matches between each $\alpha$ and \emph{at most one} $\alpha$ to its right.
For example, in the above case of $\alpha_1, \dots, \alpha_k$, when the algorithm checks if a match where $e$ matches $\alpha_1$ exists, it only examines the matches between $\alpha_1$ and $\alpha_k$.
Moreover, the algorithm makes only one pass from $\alpha_1$ to $\alpha_k$ to check all the necessary matches.
The following definition and lemma explain why this is correct.

\begin{definition2} \label{def:fwd}
    For every $i \in \myarr{Idx}_\alpha$, define $f(i) \defeq \max \{ j \in \myarr{Idx}_\alpha \mid j \leq i + |\alpha| - 1 \}$.  
\end{definition2}

\begin{lemma} \label{lem:reduce}
    For any positions $i,j \in \myarr{Idx}_\alpha$ such that $i < j < f(i)$,
    neither $w[i .. j-1]$ nor $w[j .. f(i)-1]$, nor any of their prefixes, is $\alpha$-extendable.
\end{lemma}
Therefore, $\algmatchiiib(\alpha)$ only needs to examine matches between each $\alpha$ and the rightmost $\alpha$ it overlaps.  
Moreover, when checking each $\alpha$ at position $j$ between $i$ and $f(i)$, the algorithm can skip the steps from $j$ to $f(i)-1$ and start from $f(i)$.
Thus, the overall checks can be done in $\asympO{nm_{e_1}}$ time using NFA simulation with oracles $\myarr{Pre},\myarr{Suf},\myarr{Pre}_\alpha$ and injection.
Recall that $\myarr{Pre}[i] = \mytrue \iff w[..i] \in L(e_0)$ for $i \in \myint{0}{n}$, $\myarr{Suf}[j] = \mytrue \iff w[j..] \in L(e_2)$ for $j \in \myint{1}{n+1}$ and $\myarr{Pre}_\alpha[k] = \mytrue \iff \alpha[..k] \in L(e)$ for $k \in \myint{1}{|\alpha|}$.

We explain in detail how $\algmatchiiib$ works. \cref{alg:match3b} shows the pseudocode.
First, $\algmatchiiib$ computes the array $\myarr{Fwd}$ which represents $f$ in $\asympO{n}$ time (lines~\ref{alg:match3b:belbegin} to \ref{alg:match3b:belend}).  It uses two pointers $\jleft,\jright$ and updates them so that the invariant $\myarr{Fwd}[j_1] = f(\myarr{Idx}_\alpha[j_1])$ holds every time line~\ref{alg:match3b:belasgn} is executed.  
Then, the algorithm scans each position $\inext$ of $\myarr{Idx}_\alpha$ with the starting position $\fnext$ of the rightmost $\alpha$ which the $\alpha$ at $\inext$ overlaps until the guard of the if statement at line~\ref{alg:match3b:valleft} becomes true.  
The guard has the following purpose.
In the if statement, the algorithm will check a match between a prefix of the $\alpha$ at $\inext$ and a prefix of that at $\fnext$.  Prior to this, the guard excludes the cases (1) $\inext = \fnext$ and (2) $e_0$ does not match the prefix to the left of the $\alpha$ at $\inext$.
It also excludes the case (3) $\fnext = \fprev$ to skip unnecessary checks.

If the guard holds, then the algorithm executes lines~\ref{alg:match3b:reset} to \ref{alg:match3b:test}.
The for loop in lines~\ref{alg:match3b:for} to \ref{alg:match3b:nfasim} is similar to that of $\algintmed$ (lines~\ref{alg:intmed:for} to \ref{alg:intmed:nfasim} of \cref{alg:intmed}).  
Each step performs injection and at line~\ref{alg:match3b:test} the algorithm checks the existence of a match whose backreferenced substring is the prefix of $\alpha$ which starts at $\inext$ and ends at $i$ between the $\alpha$ at $\inext$ and the one at $\fnext$.
Note that the length of the prefix is $i - \inext + 1$. 
The injection is performed only if $e$ matches the prefix and $e_2$ matches the remaining suffix $w[\fnext + i - \inext ..]$.
The for loop starts with $i = \max\{ \inext, \fprev \}$ because the checks for the prefixes of $w[\inext .. \fprev-1]$ can be skipped when $\inext < \fprev$, as mentioned in the paragraph following \cref{lem:reduce}.
This and condition (3) above ensure the linear time complexity of the algorithm with respect to $n$.

We show the correctness of $\algmatchiiib$.  The following is similar to \cref{prop:match2simset}.

\begin{proposition} \label{prop:match3bsimset}
    Let $i_1 < i_2 < i_3 < \cdots$ be the positions in $\myarr{Idx}_\alpha$.
    Every time line~\ref{alg:match3b:test} is reached at an iteration with $\inext = i_j$, we have $S = \bigcup_{i \in I} \Delta(\myecl(q_0), w[i+1 .. f(i_j)-1])$ where $I = \{ i \in \myint{\max\{i_j, f(i_{j-1})\}}{f(i_j)-1} \mid w[i_j .. i] \in L(e) \text{ and } w[f(i_j)+i-i_j+1] \in L(e_2) \}$.  Here, we regard $f(i_{j-1}) = 0$ when $j=1$.
\end{proposition}

\begin{lemma}[Correctness of $\algmatchiiib$] \label{lem:match3b}
    Let $\alpha$ be a right-maximal repeat of $w$.
    There exists a prefix $\beta$ of $\alpha$ such that $e$ matches $\beta$ and $e_0 \beta e_1 \beta e_2$ matches $w$ if $\algmatchiiib(\alpha)$ returns $\mytrue$.
    Conversely, $\algmatchiiib(\alpha)$ returns $\mytrue$ if there exists an $\alpha$-extendable prefix $\beta$ of $\alpha$ such that (i) $e$ matches $\beta$, (ii) $e_0 \beta e_1 \beta e_2$ matches $w$ and (iii) the two occurrences of $\beta$ are not $\alpha$-separable in $w$.
\end{lemma}

\begin{example} \label{ex:match3b}
    We illustrate $\algmatchiiib$ using the same instance as in \cref{ex:match2,ex:match3a}.  We consider the case where $\alpha = \mystr{abbabba}$, which is an overlapping right-maximal repeat of $w$ and whose occurrence array $\myarr{Idx}_\alpha$ is $[1,4,7]$.  
    \cref{fig:match3}(\subref{subfig:match3b}) shows part of the execution.  
    In this case, as mentioned in the paragraph immediately after \cref{lem:reduce}, the algorithm only needs to examine matches between the occurrences of $\alpha$ at positions~1 and $f(1) = 7$.  The squares at positions 1, 3 and 5 mark where injection was performed because $\mystr{a}$, $\mystr{abb}$ and $\mystr{abbab}$ are prefixes of $\alpha$ that $e$ matches and $e_2$ matches whose remaining suffix of $w$ (i.e., $\myarr{Suf}[7] = \myarr{Suf}[9] = \myarr{Suf}[11] = \mytrue$), respectively.  
    It returns $\myfalse$ because $e_1$ matches none of $w[2..6]$, $w[4..6]$ and $w[6]$.  Note that $\mystr{a}$ and $\mystr{abb}$ are non-$\alpha$-extendable prefix of $\alpha$, and the checks for these in this execution are actually redundant.
\end{example}

\section{Related Work}
\label{sec:related}

We first mention efficient solutions of the pure regular expression matching problem.  
The improvement of the $O(nm)$-time solution using NFA simulation was raised as an unsolved problem by Galil~\cite{galil1985open}, but in 1992, Myers successfully resolved it in a positive manner~\cite{myers1992four}. Since then, further improvements have been made by researchers, including Bille~\cite{bille2006new,bille2008fast,bille2009faster}. On the other hand, Backurs and Indyk have shown that under the assumption of the strong exponential time hypothesis, no solution exists within $O((nm)^{1-\epsilon})$ time for any $\epsilon > 0$~\cite{backurs2016regular}.  Recently, Bille and G{\o}rtz have shown the complexity with respect to a new parameter, the total size of the simulation sets in an NFA simulation $\sum_{i=0}^{n}{|S^{(i)}|}$, in addition to $n$ and $m$~\cite{bille2024sparse}.

We next discuss prior work on the matching problem of rewbs.
The problem can be solved by simulating \emph{memory automata} (MFA), which are a model proposed by Schmid~\cite{schmid2016characterising} with the same expressive power as rewbs.
An MFA has additional space called \emph{memory} to keep track of matched substrings.
A \emph{configuration} of an MFA $M$ with $k$ memories is a tuple $(q,u,(x_1,s_1), \dots, (x_k,s_k))$ where $q$ is a current state, $u$ is the remaining input string and $(x_j,s_j)$ ($j \in \myint{1}{k}$) is the pair of the content substring $x_j$ and the state $s_j$ of memory $j$.
Therefore, the number of configurations of $M$ equivalent to a given rewb on a given string is $\asympO{n^{2k+1}m}$. 
Because each step of an MFA simulation may involve $\asympO{n}$ character comparisons, this gives a solution to the rewb matching problem that runs in $\asympO{n^{2k+2}m}$ time.  
Davis et al.~gave an algorithm with the same time complexity as this~\cite{davis2021using}.
Furthermore, by precomputing some string indices as in this paper, a substring comparison can be done in constant time, making it possible to run in $\asympO{n^{2k+1}m}$ time.
Therefore, for the rewbs considered in this paper, these algorithms take time cubic in $n$ because $k=1$ for these rewbs, and our new algorithm substantially improves the complexity, namely, to quadratic in $n$.

Regarding research on efficient matching of rewbs, Schmid proposed the \emph{active variable degree} (avd) of MFA and discussed the complexity with respect to avd~\cite{schmid2019regular}.  Roughly, avd is the minimum number of substrings that needs to be remembered at least once per step in an MFA simulation.
For example, in a simulation of an MFA equivalent to the rewb $(a^\ast)_1 \mybs 1 (b^\ast)_2 \mybs 2$, after consuming the substring captured by $(a^\ast)_1$ in the transition which corresponds to $\mybs 1$, configurations no longer need to keep the substring.  In other words, it only needs to remember only one substring at each step of the simulation, and hence its avd is $1$.
On the other hand, $\myavd((a^\ast)_1 (b^\ast)_2 \mybs 1 \mybs 2)$ is $2$.
The avd of the rewbs considered in this paper is always $1$, but their method takes quartic (or cubic with the simple modification on MFA simulation outlined above) time for them.
Freydenberger and Schmid proposed \emph{deterministic regular expression with backreferences} and showed that the matching problem of deterministic rewbs can be solved in linear time~\cite{freydenberger2019deterministic}.
The rewbs considered in this paper are not deterministic in general (for example, $(a^\ast)_1 \mybs 1$ is not).

Next, we mention research on efficient matching of pattern languages with bounded number of repeated variables.
A \emph{pattern with variables} is a string over constant symbols and variables.  The matching problem for patterns is the problem of deciding whether a given string $w$ can be obtained from a given pattern $p$ by uniformly substituting nonempty strings of constant symbols for the variables of $p$. 
Note that, as remarked in the introduction, rewbs can be viewed as a generalization of patterns by regular expressions.

Fernau et al.~discussed the matching problem for patterns with at most $k$ repeated variables~\cite{fernau2020pattern}.  
A \emph{repeated variable} of a pattern is a variable that occurs in the pattern more than once.
In particular, for the case $k=1$, they showed the problem can be solved in quadratic time with respect to the input string length $n$.
The patterns with one repeated variable and the rewbs considered in this paper are independent.
While these patterns can use the variable more than twice, these rewbs can use regular expressions.
Therefore, our contribution has expanded the variety of languages that can be expressed within the same time complexity with respect to $n$.

The algorithm by Fernau et al.~\cite{fernau2020pattern} leverages \emph{clusters} defined over the suffix array of an input string, which are related to right-maximal repeats.  Moreover, it is similar to our approach in that it does the examination while enumerating candidate assignments to the repeated variable.
The key technical difference is that, as demonstrated in their work, matching these patterns can be reduced to finding a canonical match by dividing the pattern based on wildcard variables.
In contrast, matching the rewbs considered in this paper requires handling the general regular expression matching, particularly the substring matching of the middle expression $e_1$, which makes such a reduction not applicable even when the number of variable occurrences is restricted to $2$.

Regarding the squareness checking problem mentioned in the introduction, Main and Lorentz~\cite{main1985linear} and Crochemore~\cite{crochemore1986transducers} showed a linear-time solution on a given alphabet.  However, both solutions rely on properties specific to square-free strings, and extending them to the matching problem considered in this paper seems difficult.

Finally, we mention related work on the techniques and concepts from automata theory and stringology used in this paper.
A similar approach to using oracles such as $\myarr{Pre}$ and $\myarr{Suf}$ in NFA simulation is used in research on efficient matching of regular expressions with lookarounds~\cite{mamouras2024efficient,barriere2024linear}.
Summarization has been applied to parallel computing for pure regular expression matching~\cite{ladner1980parallel,hillis1986data,sinya2013sfa}.
Regarding injection, NFA simulation itself uses injection internally to handle concatenation of regular expressions.  That is, an NFA simulation of $e_1 e_2$ can be seen as that of $e_2$ that injects the $\varepsilon$-closure of the initial state of the NFA $N_{e_2}$ whenever $e_1$ matches the input string read so far.  Nonetheless, our use of these automata-theoretic techniques for efficient matching of rewbs is novel.
In fact, to our knowledge, this paper is the first to propose an algorithm that combines these techniques.

The right-maximal repeats of a string are known to correspond to the internal nodes of the \emph{suffix tree} of the string~\cite{gusfield1997book}.  Kasai et al.~first introduced a linear-time algorithm for traversing the internal nodes of a suffix tree using a suffix array~\cite{kasai2001linear}. 
Subsequently, Abouelhoda et al.~introduced the concept of the \emph{LCP-interval tree} to make their traversal more complete~\cite{abouelhoda2004replacing}.  As stated in \cref{sec:prelim}, our $\algenumrm$ that enumerates the right-maximal repeats with the sorted starting positions of their occurrences is based on their algorithm.
To our knowledge, our work is the first to apply these stringology concepts and techniques to efficient matching of rewbs.

\section{Conclusion}
\label{sec:conc}

In this paper, we proposed an efficient matching algorithm for rewbs of the form $e_0 (e)_1 e_1 \backslash 1 e_2$ where $e_0, e, e_1, e_2$ are pure regular expressions, which are fundamental and frequently used in practical applications.  
As stated in the introduction and \cref{sec:related}, it runs in $\asympO{n^2m^2}$ time, improving the best-known time complexity for these rewbs when $n > m$.  Because $n$ is typically much larger than $m$, this is a substantial improvement.

Our algorithm combines ideas from both stringology and automata theory in a novel way.  The core of our algorithm consists of two techniques from automata theory, injection and summarization.  Together, they enable the algorithm to do all the examination for a fixed right-maximal repeat and its extendable prefixes, which are concepts from stringology, instead of examining each individually.  By further leveraging a subtle property of extendable prefixes, our algorithm correctly solves the matching problem in time quadratic in $n$.

A possible direction for future work is to further reduce the time complexity of the algorithm.
A natural next step would be to use \emph{maximal repeats} instead of right-maximal repeats. While this would not change the worst-case complexity with respect to $n$~\cite{crochemore1997direct,raffinot2001maximal}, it could lead to faster performance for many input strings.
Another possible direction is to extend the algorithm to support more general rewbs as mentioned in \cref{rem:rewb}.
The extension of our algorithm with support for other practical extensions such as lookarounds is also challenging.

\bibliography{refs}

\ifFULL
\appendix
\section{Algorithm for Enumerating Right-Maximal Repeats}
\label{app:enum}

First, we review \emph{suffix arrays}.
We assume that the alphabet $\Sigma$ is totally ordered and has the smallest character \mystr{\$}.  Let $w$ be a string of length $n$ having $\mystr{\$}$ at the end and nowhere else.
The \emph{suffix array} $\myarr{SA}$ of $w$ is defined as the lexicographically ordered array of all the suffixes of $w$.
More precisely, $\myarr{SA}$ is the permutation of $\{1, \dots, n \}$ such that $w[\myarr{SA}[1]..] \mylexlt \cdots \mylexlt w[\myarr{SA}[n]..]$ where $\mylexlt$ denotes the lexicographical order.
Suffix arrays are often used with additional data structures, such as LCP-arrays.  The \emph{LCP-array} $\myarr{LCP}$ is the array whose $i$-th element is the length of the longest common prefix of the suffixes $w[\myarr{SA}[i-1]..]$ and $w[\myarr{SA}[i]..]$.  
It is well known that both $\myarr{SA}$ and $\myarr{LCP}$ can be constructed in linear time (refer to Ohlebusch~\cite{ohlebusch2013bioalg} or Louza et al.~\cite{louza2020construction}).
For example, $\myarr{SA}$ and $\myarr{LCP}$ of \mystr{mississimiss\$} are shown on the left side of \cref{fig:satblint}.

\begin{figure}[t] 
    \centering
    \includegraphics[scale=0.63]{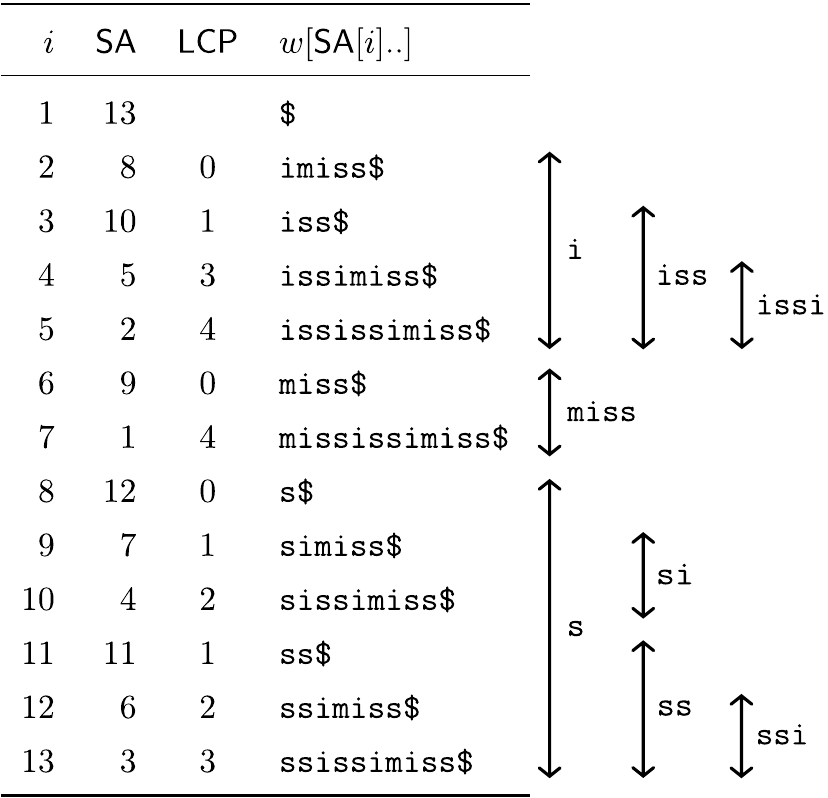}
    \caption{$\myarr{SA}$ and $\myarr{LCP}$ of \mystr{mississimiss\$} (left), and its LCP-intervals (right).}
    \label{fig:satblint}
\end{figure} 

Using these, we can enumerate all right-maximal repeats $\alpha$ of $w$ with the sorted array $\myarr{Idx}_\alpha$ of the starting positions of the occurrences of $\alpha$ in $\asympO{n^2}$ time, as we will explain below.
Right-maximal repeats are known to have a one-to-one correspondence with the internal nodes of the suffix tree~\cite{gusfield1997book}, which have a one-to-one correspondence with the concept called LCP-intervals, introduced by Abouelhoda et al.~\cite{abouelhoda2004replacing}.  An \emph{LCP-interval} is intuitively an index interval $[u,v]$ of $\myarr{SA}$ that cannot be extended without changing the longest common prefix of its corresponding suffixes.  We call the length of the longest common prefix \emph{LCP-length}. 
The right side of \cref{fig:satblint} shows an example of LCP-intervals.
%



\begin{algorithm}[t]
    \caption{$\algenumrm$ (adapted from Algorithm~4.4 in \cite{abouelhoda2004replacing}) \label{alg:enumrm}}
    \DontPrintSemicolon
    \SetKwInOut{Input}{Input}\SetKwInOut{Output}{Correctness}
    \SetKw{KwNot}{not}
    $push(\langle 0, \mynil \rangle); \;\; \myarr{LCP}[n+1] \leftarrow 0$\;
    \For{$i \leftarrow 2$ \KwTo $n$}{
        \lIf{$\myarr{LCP}[i+1] > top.lcp$}{
            $push(\langle \myarr{LCP}[i+1], \{\myarr{SA}[i]\})$
        }
        \uElseIf{$\myarr{LCP}[i+1] = top.lcp$}{
            \lIf{$top.lcp \neq 0$}{insert $\myarr{SA}[i]$ into $top.\myarr{Idx}$}
        }
        \Else{
            insert $\myarr{SA}[i]$ into $top.\myarr{Idx}$\;
            \While{$\myarr{LCP}[i+1] < top.lcp$}{
                $rmrep \leftarrow pop; \;\; process(rmrep)$\;
                \uIf{$\myarr{LCP}[i+1] \leq top.lcp$}{
                    \lIf{$top.lcp \neq 0$}{merge $rmrep.\myarr{Idx}$ into $top.\myarr{Idx}$}
                }
                \lElse{
                    $push(\langle \myarr{LCP}[i+1], rmrep.\myarr{Idx} \rangle)$
                }
            }
        }
    }
\end{algorithm}

We show the enumeration subroutine used in our algorithm.
\cite{abouelhoda2004replacing} showed an $\asympO{n}$-time enumeration algorithm for right-maximal repeats.
By slightly modifying this, we obtain an $\asympO{n^2}$-time algorithm $\algenumrm$ that enumerates each right-maximal repeat $\alpha$ with the sorted array $\myarr{Idx}_\alpha$ of all starting positions of the occurrences of $\alpha$.  

We explain in detail how the algorithm works.  \cref{alg:enumrm} shows the pseudocode.
The algorithm uses a stack to manage the visited LCP-intervals.
An LCP-interval $I$ being visited is represented by the pair $\langle lcp, \myarr{Idx} \rangle$ consisting of the LCP-length $lcp$ of $I$ and the array $\myarr{Idx}$ representing $\{ \myarr{SA}[j] \mid j \in I \text{ and } j \leq i \}$.
%
Initially, the stack has $\langle 0, \mynil \rangle$. 
Next, the algorithm repeats the following steps for $i = 2, 3, \dots, n$.
Let $top = \langle lcp, \myarr{Idx} \rangle$ denote the pair at the top of the stack.
We write $I_{top}$ for the LCP-interval represented by $top$.

In each iteration of the for loop, the algorithm first compares $lcp$ with $\myarr{LCP}[i+1]$.  We assume $\myarr{LCP}[n+1] = 0$.
(1) If $\myarr{LCP}[i+1] > lcp$, it pushes $\langle \myarr{LCP}[i+1], \{ \myarr{SA}[i] \} \rangle$ into the stack, because index $i$ is at the left end of an LCP-interval whose LCP-length is greater than $lcp$.
(2) If $\myarr{LCP}[i+1] = lcp$ and the stack has no LCP-interval, it inserts $\myarr{SA}[i]$ into $\myarr{Idx}$ while preserving the ascending order, because index $i$ belongs to the LCP-interval $I_{top}$.
(3) If $\myarr{LCP}[i+1] < lcp$, it first inserts $\myarr{SA}[i]$ into $\myarr{Idx}$ because index $i$ is at the right end of $I_{top}$. Then, it pops all LCP-intervals whose right end is at $i$ while passing each popped entry $rmrep$ to a placeholder function $process$.  A subtle point lies in the relationship between the popped LCP-interval and the LCP-interval that was directly beneath it on the stack (see Theorem~4.3 of \cite{abouelhoda2004replacing} for details).  Depending on this relationship, $\myarr{Idx}$ of $rmrep$ must be merged while preserving the ascending order or pushed as a part of a new LCP-interval accordingly.

For example, the algorithm enumerates the right-maximal repeats of \mystr{mississimiss\$} in the order $\langle 4, [2,5] \rangle \to \langle 3, [2,5,10] \rangle \to \langle 1, [2,5,8,10] \rangle \to \langle 4, [1,9] \rangle \to \langle 2, [4,7] \rangle \to \langle 3, [3,6] \rangle \to \langle 2, [3,6,11] \rangle \to \langle 1, [3,4,6,7,11,12] \rangle$ (see \cref{fig:satblint}). 
The algorithm runs in $\asympO{n^2}$ time because each step of (1)(2)(3) takes $\asympO{n}$ time, and each is executed at most $\asympO{n}$ times.  The space complexity is $\asympO{n}$.

\section{Omitted Proofs}
\label{app:pf}

\begin{proof}[Proof of \cref{lem:match}]
    Immediate from \cref{lem:match3a,lem:match3b}.
\end{proof}

\begin{proof}[Proof of \cref{lem:injecting}]
    Immediate from the definition of $\Delta$.
\end{proof}

\begin{proof}[Proof of \cref{prop:match1simset}]
    It suffices to prove the statement of the proposition with ``line~\ref{alg:match1:l:test}'' replaced by ``line~\ref{alg:match1:l:update}.''
	We prove by induction on $j$.
    We write $J'_j$ for $\{ j' \in \myint{1}{|\myarr{Idx}_\alpha|} \mid i_{j'} + |\alpha| \le i_j \text{ and } w[..i_{j'}-1] \in L(e_0) \}$ and $S_j$ for $\bigcup_{j' \in J'_j} \Delta(\myecl(q_0), w[i_{j'}+|\alpha|, i_{j}-1])$.
    We show $S = S_j$ right after line~\ref{alg:match1:l:rett} in the iteration with $\inext = i_j$.
	\begin{itemize}
        \item Case $j=1$:  Because $\ique = \mynil$, the if statement in lines~\ref{alg:match1:l:ifiqueisnotnil} to \ref{alg:match1:l:rett} is skipped and $S = \emptyset$ holds when it reaches line~\ref{alg:match1:l:update}.  On the other hand, $J'_1 = \emptyset$ because no $\alpha$ occurs before the $\alpha$ at $i_1$.  Therefore, $S = S_1 = \emptyset$.
        \item Case $j-1 \to j$: By the induction hypothesis, we have $J'_{j-1} = \{ j' \in \myint{1}{|\myarr{Idx}_\alpha|} \mid i_{j'} + |\alpha| \le i_{j-1} \text{ and } w[..i_{j'}-1] \in L(e_0) \}$ and $S = S_{j-1}$ when it reaches line~\ref{alg:match1:l:update} in the iteration with $\inext = i_{j-1}$.
            At line~\ref{alg:match1:l:que}, $\iprev = i_{j-1}$ holds.
            Let $J'_{j-1,j}$ denote $\{ j' \mid i_{j-1} \leq i_{j'} + |\alpha| < i_j \text{ and } w[..i_{j'}-1] \in L(e_0) \}$.
            Observe that $J'_j = J'_{j-1} \cup J'_{j-1,j}$.
            From the assumption that $\alpha$ is a nonoverlapping repeat, $J'_{j-1,j}$ is either $\emptyset$ or $\{j-1\}$.
            We consider two cases.
            \begin{itemize}
                \item Case $\myarr{Pre}[\iprev-1] = \mytrue$:  In this case, $e_0$ matches $w[..i_{j-1}-1]$.  Therefore, $J'_{j-1,j} = \{j-1\}$.
                    At line~\ref{alg:match1:l:que}, $\ique$ becomes $i_{j-1} + |\alpha| - 1$.
                    Then, $\inext$ becomes $i_j$ at line~\ref{alg:match1:l:foridx}.  Now, $\ique \neq \mynil$ and the algorithm enters the for loop at line~\ref{alg:match1:l:foraltoal}.  We further divide the case into two. 
                    \begin{itemize}
                        \item Case $S = \emptyset$: In this case, we have $J'_{j-1} = \emptyset$.
                            Lines~\ref{alg:match1:l:nfasim} and \ref{alg:match1:l:inj} are skipped until $i$ becomes $\ique = i_{j-1} + |\alpha| - 1$.  When $i = \ique$, the algorithm injects $\myecl(q_0)$ into $S$ at line~\ref{alg:match1:l:inj} and starts an NFA simulation on $w[i_{j-1}+|\alpha| .. i_j-1]$.  Thus, right after line~\ref{alg:match1:l:rett}, $S = \Delta(\myecl(q_0), w[i_{j-1}+|\alpha| .. i_j-1])$ holds.  Therefore, $J'_j = J'_{j-1,j} = \{j-1\}$ and $S = S_j$.
                        \item Case $S \neq \emptyset$:  After the for loop at lines~\ref{alg:match1:l:foraltoal} to \ref{alg:match1:l:inj} is executed, $S$ becomes 
                            \[
                                \Delta(\Delta(S_{j-1},w[i_{j-1} .. i_{j-1}+|\alpha|-1]) \cup \myecl(q_0), w[i_{j-1}+|\alpha| .. i_j - 1]).  
                            \]
                            By the induction hypothesis, 
                                $S = \Delta(\Delta(\bigcup_{j'\in J'_{j-1}}\Delta(\myecl(q_0),w[i_{j'}+|\alpha| .. i_{j-1}-1]), w[i_{j-1} .. i_{j-1}+|\alpha|-1])\cup\myecl(q_0), w[i_{j-1}+|\alpha| .. i_j-1])$.
By \cref{lem:injecting}, this is equal to
                                $\bigcup_{j' \in J'_{j-1} \cup \{j-1\}} \Delta(\myecl(q_0),w[i_{j'}+|\alpha| .. i_j-1])$.  We have $J'_j = J'_{j-1} \cup \{j-1\}$.
                            Therefore, $S = S_j$.
                    \end{itemize}
                \item Case $\myarr{Pre}[\iprev-1] = \myfalse$:  In this case, $e_0$ does not match $w[..i_{j-1}-1]$ and $J'_{j-1,j} = \emptyset$.  At line~\ref{alg:match1:l:foridx}, $\inext$ becomes $i_j$.  We further divide the case into two.
                    \begin{itemize}
                        \item Case $\ique = \mynil$:  In this case, we have $S = \emptyset$.  The if statement at lines~\ref{alg:match1:l:ifiqueisnotnil} to \ref{alg:match1:l:rett} is skipped and $S = S_{j-1}$ holds right after line~\ref{alg:match1:l:rett}.  Because $J'_j = J'_{j-1} = \emptyset$, we have $S = S_j$.
                        \item Case $\ique \neq \mynil$:  After the for loop at lines~\ref{alg:match1:l:foraltoal} to \ref{alg:match1:l:inj} is executed, $S$ becomes 
                            \[
                                \Delta(S_{j-1}, w[i_{j-1} .. i_j-1]).
                            \]By \cref{lem:injecting} and the induction hypothesis, $S = \Delta(\bigcup_{j'\in J'_{j-1}} \Delta(\myecl(q_0), w[i_{j'}+|\alpha| .. i_{j-1}-1]), w[i_{j-1} .. i_j-1]) = \bigcup_{j' \in J'_{j-1}} \Delta(\myecl(q_0), w[i_{j'}+|\alpha| .. i_j-1])$ holds.  Because $J'_j = J'_{j-1}$, we have $S = S_j$.
                    \end{itemize}
            \end{itemize}
	\end{itemize}
\end{proof}

\begin{proof}[Proof of \cref{lem:match1}]
    Let $i_1 < i_2 < \cdots$ be the positions in $\myarr{Idx}_\alpha$.
    Suppose that $\algmatchi(\alpha)$ returns $\mytrue$.
    From the definition of the algorithm and by \cref{prop:match1simset}, $\alpha \in L(e)$ and there exist $i_{j'}, i_j \in \myarr{Idx}_\alpha$ such that (1) $w[..i_{j'}-1] \in L(e_0)$, (2) $i_{j'}+|\alpha| < i_j$, (3) $\Delta(\myecl(q_0), w[i_{j'}+|\alpha| .. i_j-1]) \cap F \neq \emptyset$ and (4) $w[i_j+|\alpha|..] \in L(e_2)$.  Therefore, $e$ matches $\alpha$ and $r_\alpha$ matches $w$.
    The other direction follows in the same manner.
\end{proof}

\begin{proof}[Proof of \cref{prop:match2simset}]
    The proof follows similarly to \cref{prop:match1simset}.
\end{proof}

\begin{proof}[Proof of \cref{lem:intmed}]
    Recall that $I = \{ i \in \myint{\ibeg}{\iend} \mid w[\iend - |\alpha| + 1 .. i] \in L(e) \text{ and } w[i+1 ..] \in L(e_2) \}$.
    For every $i \in \myint{\ibeg}{\iend}$, we write $I_{\le i}$ for $\{ i' \in I \mid i' \le i \}$.
        Because the statement easily holds when $\ibeg = \iend$, we assume $\ibeg \le \iend-1$ in what follows.
    We prove the following statement by induction:
    
    \begin{claim*}
        Suppose that $\ibeg \le \iend-1$.
        Every time line~\ref{alg:intmed:nfasim} has been executed at an iteration with $i \in \myint{\ibeg}{\iend-1}$, we have $T = \bigcup_{i'\in I_{\le i}} \Delta(\myecl(q_0), w[i'+1 .. i+1])$.
    \end{claim*}
    \begin{claimproof}
        The base case when $i = \ibeg$ is obvious.  
        Suppose that the step $i-1$ of the for loop has finished, and line~\ref{alg:intmed:nfasim} has been executed.  By the induction hypothesis, $T = \bigcup_{i' \in I_{\le i-1}} \Delta(\myecl(q_0),w[i'+1 .. i])$.  In step $i$, we consider two cases.
        \begin{itemize}
            \item Case $i \in I$:  Observe that $I_{\le i} = I_{\le i-1} \cup \{i\}$.  In this case, the algorithm injects $\myecl(q_0)$ into $T$ at line~\ref{alg:intmed:inj}.  Then, immediately after line~\ref{alg:intmed:nfasim} has been executed, $T$ becomes $\Delta(\bigcup_{i' \in I_{\le i-1}} \Delta(\myecl(q_0),w[i'+1 .. i]) \cup \myecl(q_0), w[i+1])$.  By \cref{lem:injecting}, this is equal to $\bigcup_{i'\in I_{\le i}} \Delta(\myecl(q_0), w[i'+1 .. i+1])$.
            \item Case $i \notin I$:  Observe that $I_{\le i} = I_{\le i-1}$.  In this case, the if statement at line~\ref{alg:intmed:inj} is skipped.  If $T = \emptyset$ right after line~\ref{alg:intmed:inj}, then $I_{\le i-1}$ is also empty, which implies $I_{\le i} = \emptyset$, and thus the statement holds.  Otherwise, immediately after line~\ref{alg:intmed:nfasim} has been executed, $T$ becomes $\Delta(\bigcup_{i' \in I_{\le i-1}} \Delta(\myecl(q_0),w[i'+1 .. i]), w[i+1])$.  By \cref{lem:injecting}, this is equal to $\bigcup_{i'\in I_{\le i}} \Delta(\myecl(q_0), w[i'+1 .. i+1])$.
        \end{itemize}
    \end{claimproof}

    Finally, we divide the cases into two for the last iteration of the for loop.
    \begin{itemize}
        \item Case $\iend \in I$: after line~\ref{alg:intmed:inj} is executed, $T$ becomes $\bigcup_{i'\in I_{\le \iend-1}} \Delta(\myecl(q_0), w[i'+1 .. \iend]) \cup \myecl(q_0) = \bigcup_{i'\in I_{\le \iend}} \Delta(\myecl(q_0), w[i'+1 .. \iend])$.
        \item Case $\iend \notin I$: similarly to the above case.
    \end{itemize}
\end{proof}

\begin{proof}[Proof of \cref{lem:match2}]
Let $i_1 < i_2 < \cdots$ be the positions in $\myarr{Idx}_\alpha$.
Suppose that $\algmatchii(\alpha)$ returns $\mytrue$ in the iteration with $\inext = i_j$.
    From the definition of $\algmatchii$ and by \cref{prop:match2simset} and \cref{lem:intmed}, there exist (1) $i \in \myint{i_j}{i_j+|\alpha|-1}$ such that $w[i_j .. i] \in L(e)$ and $w[i+1 ..] \in L(e_2)$, (2) $q_l \in \Delta(\myecl(q_0), w[i+1 .. i_j + |\alpha| - 1])$ and (3) $i_{j'}$ such that $i_{j'} + |\alpha| \le i_j$, $w[..i_{j'}-1] \in L(e_0)$ and $\Delta(\{q_l\}, w[i_{j'}+|\alpha| .. i_j - 1]) \cap F \neq \emptyset$.
    Let $k$ be $i - i_j + 1$.  Note that $w[i+1 .. i_j + |\alpha| - 1] = \alpha[k+1 ..]$.

    Let $\beta$ be the prefix $\alpha[..k]$.  From (1) above, $e$ matches $\beta$.  We claim that $e_0 \beta e_1 \beta e_2$ matches $w$, where the two $\beta$'s correspond to the ones at positions $i_{j'}$ and $i_{j}$.
    In fact, because $q_l \in \Delta(\myecl(q_0), \alpha[k+1..])$ and $\Delta(\{q_l\}, w[i_{j'}+|\alpha| .. i_j - 1]) \cap F \neq \emptyset$, it follows that $e_1$ matches $\alpha[k+1 .. ]w[i_{j'}+|\alpha| .. i_j - 1]$.  Observe that $\alpha[k+1 .. ] = w[i_{j'}+k .. i_{j'}+|\alpha|-1]$.  From this, together with (1) and (3), there is a match where $e_0$ matches $w[..i_{j'}-1]$, $e_1$ matches $w[i_{j'}+k .. i_j-1]$ and $e_2$ matches $w[i_j + k ..]$.

    Conversely, let $\beta$ be an $\alpha$-extendable prefix $\alpha[..k]$ and suppose that there is a match with $\beta$.  Because $\myrimp{\beta} = \alpha$ and $\alpha$ is nonoverlapping, the starting positions of the two $\beta$'s can be taken as $i_{j'}$ and $i_j$ satisfying $i_{j'} + |\alpha| \le i_j$.  Let $\gamma$ be $\alpha[k+1 ..]$.  Because $e_1$ matches $\gamma (w[i_{j'}+|\alpha| .. i_j - 1])$, there exists a state $q_l$ such that $q_l \in \Delta(\myecl(q_0),\gamma)$ and $\Delta(\{q_l\}, w[i_{j'}+|\alpha| .. i_j-1]) \cap F \neq \emptyset$.
    We first show that $q_l$ belongs to $T$ in the iteration with $\inext = i_j$.
    Observe that $T = \algintmed(i_j,i_j+|\alpha|-1)$ and $\gamma = w[i_j + |\beta| .. i_j + |\alpha| - 1]$.
    By \cref{lem:intmed}, it suffices to show that $i_j+|\beta|-1$ belongs to the index set $I$ where $I$ defined in the lemma, and this can be easily checked.
    Next, we show that $\mathcal{S}[l] \cap F \neq \emptyset$ in the same iteration.
    By \cref{prop:match2simset}, it suffices to show that $i_{j'}$ belongs to the index set $J'$ where $J'$ defined in the proposition, and this can also be easily checked.
    Therefore, $\algmatchii$ returns $\mytrue$ in the iteration with $\inext = i_j$.
\end{proof}

\begin{proof}[Proof of \cref{lem:once}]
    Assume to the contrary that $\beta$ is $\alpha$-extendable.
    If $\alpha$ contains an occurrence of $\beta$ at any position other than the beginning, the occurrence of $\beta$ could extend to another occurrence of $\alpha$, which contains another occurrence of $\beta$.  By repeating this infinitely, it could extend to the right without bound.
\end{proof}

\begin{proof}[Proof of \cref{lem:match3a}]
    Let $i_1 < i_2 < \cdots$ be the positions in $\myarr{Idx}_\alpha$.
    \begin{claim*}
        Suppose that $\myarr{Que} \neq \mynil$ holds for the first time in an iteration with $\inext = i_{j_0}$.  For each $i_j \ge i_{j_0}$, right after executing the iteration with $\inext = i_j$, $\myarr{Que}$ represents $\{ i_{j'}+|\alpha|-1 \mid i_{j'} \le i _j \le i_{j'} + |\alpha| - 1 \land w[.. i_{j'}-1] \in L(e_0) \}$ in sorted order.
    \end{claim*}
    \begin{claimproof}
        We prove by induction on $j$.  The base case $j = j_0$ easily holds.
        At the beginning of the iteration with $\inext = i_j$, $\myarr{Que}$ represents $\{ i_{j'}+|\alpha|-1 \mid i_{j'} \le i _{j-1} \le i_{j'} + |\alpha| - 1 \land w[.. i_{j'}-1] \in L(e_0) \}$.  Then, right after completing the execution of the for loop in lines~\ref{alg:match3a:iptoin} to \ref{alg:match3a:dequeue}, $\myarr{Que}$ becomes $\{ i_{j'}+|\alpha|-1 \mid i_{j'} \le i _{j-1} \land i_j \le i_{j'} + |\alpha| - 1 \land w[.. i_{j'}-1] \in L(e_0) \}$.  We divide the case into two.
        \begin{itemize}
            \item Case $w[.. i_j - 1] \in L(e_0)$: The algorithm enqueues $i_j + |\alpha| - 1$ into $\myarr{Que}$ at line~\ref{alg:match3a:enqueue}.  Therefore, the claim holds for $i_j$.
            \item Case $w[.. i_j - 1] \notin L(e_0)$: The if statement in lines~\ref{alg:match3a:preiprev} to \ref{alg:match3a:enqueue} is skipped.  Therefore, the claim holds for $i_j$.
        \end{itemize}
    \end{claimproof}

    \begin{claim*}
        Every time line~\ref{alg:match3a:test} is reached at iteration with $\inext = i_j$, we have $\mathcal{S}[l] = \bigcup_{j' \in J'} \Delta(q_l, w[i_{j'}+|\alpha|, i_{j}-1])$ where $J' = \{ j' \in \myint{1}{|\myarr{Idx}_\alpha|} \mid i_{j'} + |\alpha| \le i_j \text{ and } w[..i_{j'}-1] \in L(e_0) \}$.
    \end{claim*}
    \begin{claimproof}
        With the above claim, the proof follows in the same way as for \cref{prop:match1simset}.
    \end{claimproof}

    The lemma can be proved in the same way as \cref{lem:match2}, using the above claim and \cref{lem:intmed}.
\end{proof}

\begin{proof}[Proof of \cref{lem:reduce}]
    The $\alpha$ at position $j$ contains $w[f(i) .. j+|\alpha|-1]$, which is a longer prefix than $w[i .. j-1]$, twice.
    Also, the $\alpha$ at position $i$ contains $w[j .. f(i)-1]$ twice.
    The lemma follows from these and \cref{lem:once}.
\end{proof}

\begin{proof}[Proof of \cref{prop:match3bsimset}]
 The proof follows similarly to \cref{lem:intmed}.       
\end{proof}

\begin{proof}[Proof of \cref{lem:match3b}]
    We prove only one direction.  
    Let $i_1 < i_2 < \cdots$ be the positions in $\myarr{Idx}_\alpha$.
    Suppose that there is a prefix $\beta = \alpha[..k]$ of $\alpha$ and positions $i_{j}$ and $i_{j'}$ such that (1) $\myrimp{\beta} = \alpha$, (2) $\beta \in L(e)$, (3) $i_{j} + k \le i_{j'}$, (4) $w[.. i_{j}-1] \in L(e_0)$, (5) $w[i_{j}+k .. i_{j'}-1] \in L(e_1)$, (6) $w[i_{j'}+k ..] \in L(e_2)$ and (7) the $\beta$'s are not $\alpha$-separable, i.e., $i_{j'} \le i_j + |\alpha| - 1$.
    
    We claim that $i_{j'} = f(i_j)$.  Otherwise, $i_j < i_{j'} < f(i_j)$ holds, and by \cref{lem:reduce}, $\myrimp{w[i_j .. i_{j'}-1]} \neq \alpha$, which contradicts (1) and (3).  
    In what follows, we show that the algorithm returns $\mytrue$ at an iteration with $\inext = i_j$.  We first check the guard condition at line~\ref{alg:match3b:valleft}.  Observe that $\fnext = f(i_j)$.
    Among the three conditions, the first two clearly hold.
    In fact, the third condition also holds: when $j = 1$, we have $\fprev = 0$ and the condition trivially holds.  
    Assume to the contrary that it does not hold for $j \ge 2$.  In this case, we have $\fprev = f(i_{j-1})$ and $f(i_{j-1}) = f(i_j)$.  Then, the occurrence of $\alpha$ at $i_{j-1}$ contains $\beta = w[i_j .. i_j +k-1] = w[i_{j-1} .. i_{j-1}+k-1]$ twice, violating \cref{lem:once}.
    
    By \cref{prop:match3bsimset}, at line~\ref{alg:match3b:test}, $S$ becomes $\bigcup_{i \in I} \Delta(\myecl(q_0), w[i+1 .. f(i_j)-1])$ where $I = \{ i \in \myint{\max\{i_j, f(i_{j-1})\}}{f(i_j)-1} \mid w[i_j .. i] \in L(e) \text{ and } w[f(i_j)+i-i_j+1] \in L(e_2) \}$.  Here, we regard $f(i_{j-1}) = 0$ when $j=1$.
    It suffices to show that $i_j + k - 1 \ge \max\{ i_j, f(i_{j-1})\}$.  If $i_j \ge f(i_{j-1})$, it trivially holds, so we may assume $j \ge 2$ and $i_j < f(i_{j-1})$.  
    Suppose that $f(i_{j-1}) > i_j + k - 1$.
    Let $\beta'$ be $w[i_j .. f(i_{j-1}) - 1]$, which is equal to $\beta$ or longer.
    By \cref{lem:reduce}, $\myrimp{\beta'} \neq \alpha$ holds.  This is a contradiction. 
\end{proof}

\fi

\end{document}